\documentstyle[12pt,psfig]{article}
\input xy
\xyoption{all}

\begin{document}

\hfill hep-th/0502053

\vspace{0.5in}

\begin{center}

{\large\bf GLSM's for Gerbes } \\

{\large\bf (and other toric stacks)}

\vspace{0.2in}

Tony Pantev$^1$ and Eric Sharpe$^2$ \\
$^1$ Department of Mathematics \\
University of Pennsylvania \\
David Rittenhouse Lab. \\
209 South 33rd Street \\
Philadelphia, PA  19104-6395 \\
$^2$ Departments of Physics, Mathematics \\
University of Utah \\
Salt Lake City, UT  84112 \\
{\tt tpantev@math.upenn.edu}, {\tt ersharpe@math.utah.edu} \\

\end{center}

In this paper we will discuss gauged linear sigma model
descriptions of toric stacks.  
Toric stacks have a simple description in terms of 
(symplectic, GIT) ${\bf C}^{\times}$ quotients of 
homogeneous coordinates, in exactly the same form as toric varieties.
We describe the physics of the gauged linear sigma models that
formally coincide with the mathematical description of toric stacks,
and check that physical predictions of those gauged linear sigma models
exactly match the corresponding stacks.  We also check in examples
that when a given toric stack has multiple presentations in a form
accessible as a gauged linear sigma model, that the IR physics of
those different presentations matches, so that the IR physics
is presentation-independent, making it reasonable to associate
CFT's to stacks, not just presentations of stacks.
We discuss mirror symmetry for stacks, using Morrison-Plesser-Hori-Vafa
techniques to compute mirrors explicitly, and also find a natural
generalization of Batyrev's mirror conjecture.  In the process of studying
mirror symmetry, we find some new abstract CFT's, involving fields valued
in roots of unity.

\begin{flushleft}
January 2005
\end{flushleft}

\newpage

\tableofcontents

\newpage

\section{Introduction}

String compactifications on stacks are described in \cite{tonyme}.
Briefly, every\footnote{Under extremely mild technical conditions,
essentially irrelevant for physics, as explained in \cite{tonyme}.} 
stack has a presentation as a global quotient by a 
group $[X/G]$.  A sigma model on such a presentation is simply a 
$G$-gauged sigma model on $X$.  Now, a given stack can have many presentations
of that form, yielding very different physical theories.
However, we have argued in \cite{tonyme}, and will check further here,
that the gauged sigma models defined by different presentations of a single
stack all lie in the same universality class of worldsheet RG flow.
This we take to be the central consistency condition for
the notion of ``string compactification on a stack''
to be well-defined.
In checking consistency of this hypothesis, one runs into many
potential obstacles, perhaps most importantly a mismatch between
physical deformations and stack deformations; these issues are addressed
in detail in \cite{tonyme}.

Just as toric varieties provide a nontrivial set of examples that
are easy to analyze, there is a notion of toric stacks \cite{bcs},
which have analogous properties.  Just as toric varieties can be described
by gauged linear sigma models, so too can toric stacks, and this paper
is devoted to explaining the relevant physics of such
gauged linear sigma models.

A gauged linear sigma model describing a toric stack looks very much
like a gauged linear sigma model describing a toric variety,
except that the charges with respect to some of the $U(1)$'s will
typically be nonminimal.  Now, due to peculiar properties of
two-dimensional quantum field theories,
a two-dimensional gauge theory with fields of nonminimal charges
is not the same as a two-dimensional gauge theory with fields of minimal
charges -- the two physical theories are distinguished by
nonperturbative effects, as we shall explicitly
see shortly. 

Just as the physics of the two-dimensional ${\bf C} {\bf P}^{N-1}$ model plays
an important role in understanding the physics of traditional
gauged linear sigma models, the physics of analogous models
describing gerbes over projective spaces, which we shall call the
$G {\bf P}^{N-1}$ models, play an important role in understanding
gauged linear sigma models for toric stacks, and we shall spend much
of this paper describing such models.  A gerbe, the reader should recall,
is a local orbifold by a noneffectively-acting group, and as with all
local orbifolds, can be expressed as a global quotient by a larger group.
The $G {\bf P}^{N-1}$ models we shall consider correspond to
${\bf Z}_k$ gerbes over projective spaces, meaning that they describe
local orbifolds by trivially-acting ${\bf Z}_k$'s.  The easiest
nontrivial example of such a gerbe can be expressed in a form almost
identical to the ${\bf C} {\bf P}^{N-1}$ model, except that the fields
all have charge $k$ rather than charge $1$, which is one way to describe
a local orbifold by a noneffectively-acting ${\bf Z}_k$.

We begin in section~\ref{expnonmin} by giving a general explanation for
why two-dimensional gauge theories with nonminimal charges are physically
distinct from two-dimensional gauge theories with minimal charges.
The basic reason is nonperturbative effects:  the charge dictates
which line bundle a Higgs field couples to, and different line bundles
mean different zero modes, hence different anomalies, different
correlation functions, and thus different physical theories.
We shall see many examples of this distinction throughout this paper.

In section~\ref{gpnmodels} we study examples of ${\bf Z}_k$ gerbes over
projective spaces, which we describe using gauged linear sigma
models (in which the fields have nonminimal $U(1)$ charges).
We compute physical properties of these models, such as quantum
cohomology rings.  Since we can compute quantum cohomology rings without
knowing classical cohomology rings, our computations give a very nontrivial
check of the massless spectrum conjecture presented in \cite{tonyme}.
We also are able to check presentation-independence
of universality classes, as some of these gerbes have multiple
realizations in gauged linear sigma models of rather different-looking
forms, so {\it e.g.} we can repeat calculations
in each different gauged linear sigma model, 
and check that the results are independent
of presentation.   

In section~\ref{moreexs} we briefly outline some more general examples
of toric stacks and their properties.  We briefly outline the analogue
of flops for stacks.  We also briefly discuss weighted projective stacks 
-- just like weighted projective spaces, except singularities are replaced
by local orbifold structures.  We observe that the quantum cohomology
results of \cite{ronendave2} immediately generalize to toric stacks.

In section~\ref{cygerbes} we discuss Calabi-Yau gerbes,
realized as complete intersections in toric stacks.
Physically the corresponding gauged linear sigma models are easy
to describe:  for each $U(1)$, the degree of the hypersurface must match
the sum of the charges of the chiral superfields, just as usual.
This turns out to give an easy physics derivation of a mathematical
result that all gerbes over Calabi-Yau's are themselves Calabi-Yau.
We study phases of the analogue of the quintic in a gerbe over ${\bf P}^4$,
and use the spectrum of the Landau-Ginzburg orbifold (now a noneffective
orbifold, an extension of the usual ${\bf Z}_5$ by a trivially-acting
${\bf Z}_k$) to give a very strong check of the general
massless spectrum conjecture presented in \cite{tonyme} for strings on stacks.

In section~\ref{mirrors} we study mirror symmetry for stacks.
We begin by computing Toda duals to the gerbes over projective spaces
described in section~\ref{gpnmodels}.  In addition to getting another
check of presentation-independence, by verifying that the Toda dual
depends upon the gerbe and not the presentation thereof,
we find that the Toda duals contain fields valued in roots of unity,
a result we verify by checking correlation functions in the Toda duals
and comparing to quantum cohomology results from section~\ref{gpnmodels}.
Such fields are (to the authors) rather unusual in physics,
though we also saw them from independent lines of reasoning in 
\cite{nr}, when studying marginal deformations of noneffective orbifolds.
We discuss the mirror of the gerby quintic, and gerby minimal models,
and present a conjecture for a generalization of Batyrev's mirror conjecture
to hypersurfaces in toric stacks.  Now, the data defining a toric stack
looks much like the data defining a toric variety, except that the fan
is decorated by some abelian finite-group data.
By adding fields valued in roots of unity to terms in superpotentials of
Landau-Ginzburg models, for example, we generate Newton polyhedra with
symmetric data, making an analogue of Batyrev's conjecture possible.
Note that, yet again, we see fields valued in roots of unity.

Finally, in an appendix, we review the \cite{bcs} description of
toric stacks.

\section{Two-dimensional gauge theories with nonminimal charges}
\label{expnonmin}

Since the gauged linear sigma models for many toric stacks
describe chiral superfields with nonminimal $U(1)$ charges,
let us take a few moments to explain in general terms
why nonperturbative effects distinguish such theories from two-dimensional
gauge theories with minimal charges.
Later we shall see numerous physical properties that differ
between the theories, from correlation functions to R-symmetry
anomalies, but in this section we shall just review the
basic principles.  (We would like to thank J.~Distler
and R.~Plesser for providing the detailed argument that we give
in this section.)  For a different discussion of two dimensional
gauge theories with fermions of nonminimal charges, 
see \cite{edold}[section 4].  (The discussion there is most
applicable to the present situation when $m \ll M$, in the notation
of that reference.)

To be specific, consider a gauged linear sigma model with a single
$U(1)$ gauge field, and with chiral superfields, all of charge $k$,
with $k>1$.  (Mathematically, this corresponds to a gerbe
on a projective space, as we shall review in the next section.)
One might argue that this theory should be the same as a theory
with chiral superfields of charge $1$, as follows.
Since instanton number is essentially monopole number,
from Dirac quantization since the electrons have charges a multiple
of $k$, the instantons must have charge a multiple of $1/k$,
and so zero modes of the Higgs fields in a minimal nonzero instanton
background would be sections of ${\cal O}(k/k) = {\cal O}(1)$,
just as in a minimal charge GLSM.  Making the charges nonminimal
has not changed the physics.
In order to recover the physics we have described,
we require the Higgs fields to have charge $k$ while the
instanton numbers are integral, not fractional.

Closer analysis reveals subtleties.
Let us break up the analysis into two separate cases:  first,
the case that the worldsheet is noncompact, second,
that the worldsheet is compact.  For both cases, it will be important
that the worldsheet theory is two-dimensional.

First, the noncompact case.
Since the $\theta$ angle couples to $\mbox{Tr }F$,
we can determine the instanton numbers through the periodicity of
$\theta$.  Suppose we have the physical theory described above,
namely a GLSM with Higgs fields of charge $k$,
plus two more massive fields, of charges $+1$ and $-1$.
In a two-dimensional theory, the $\theta$ angle acts as an electric
field, which can be screened by pair production, and that screening
determines the periodicity of $\theta$.
If the only objects we could pair produce were the Higgs fields
of charge $k$, then the theta angle would have periodicity
$2 \pi k$, and so the instanton numbers would be multiples
of $1/k$.  However, since the space is noncompact, and the
electric field fills the entire space, we can also pair produce
arbitrary numbers of the massive fields, which have charges
$\pm 1$, and so the $\theta$ angle has periodicity $2 \pi$,
so the instantons have integral charges.

We can phrase this more simply as follows.
In a theory with only Higgs fields of charge $k$,
the instanton numbers are multiples of $1/k$, and so the resulting
physics is equivalent to that of a GLSM with minimal charges.
However, if we add other fields of charge $\pm 1$,
then the instanton numbers are integral,
and if those fields become massive, and we work at an energy scale
below that of the masses of the fields, then we have a theory
with Higgs fields of charge $k$, and integral instanton numbers,
giving us the physics that corresponds to a gerbe target.

Thus, we see in the noncompact case that there are two
possible physical theories described by Higgs fields of charge $k$:
one is equivalent to the GLSM with minimal charges,
and the other describes the gerbe.

The analysis for the compact worldsheet case is much shorter.
Strictly speaking, to define the theory nonperturbatively on a
compact space, we must specify, by hand, the bundles that the
Higgs fields couple to.  If the gauge field is described by
a line bundle $L$, then coupling all of the Higgs fields to
$L^{\otimes k}$ is a different prescription from coupling all
of the Higgs fields to $L$.  As a result, the spectrum of zero modes
differs between the two theories, hence correlation functions and
anomalies differ between the two theories,
and so the two physical theories are very different,
as we shall see in examples later.

We shall assume throughout this paper that the worldsheet is
compact, though as we have argued the same subtlety shows up
for noncompact worldsheets.

We are interested in these physical theories
because they often crop up in describing stacks.
Since stacks are defined via their incoming maps, a precise definition
of which bundles the superfields couple to is part of the definition
of the stack, and so there is no ambiguity in which physical
theory to associate to a given (presentation of a) stack.

\section{Gerbes over projective spaces and quantum cohomology}
\label{gpnmodels}

In this section we will discuss basic properties of physical
theories describing gerbes over projective spaces,
the gerby analogue of the old ``${\bf C} {\bf P}^{N-1}$
models.''  We will occasionally refer to these gerby analogues
as $G {\bf P}^{N-1}$ models, or more specifically,
$G_a^k {\bf P}^{N-1}$ models, for a ${\bf Z}_k$ gerbe
on ${\bf P}^{N-1}$ classified by 
\begin{displaymath}
a \mbox{ mod } k \: \in \: {\bf Z}_k \: = \:
H^2\left( {\bf P}^{N-1}, {\bf Z}_k \right)
\end{displaymath}
In this notation, therefore, $G_a^k {\bf P}^{N-1} = 
G_{a + k}^k {\bf P}^{N-1}$.

\subsection{First example:  ${\bf Z}_k$ gerbe on ${\bf P}^{N-1}$}
\label{basicpngerbe}

Let us begin with an easy example of a gerbe over projective space
${\bf P}^{N-1}$.  The projective space itself is described
by a gauged linear sigma model with $N$ chiral superfields $x_i$,
each of charge one with respect to a gauged $U(1)$.
Mathematically, the projective space is the quotient
\begin{displaymath}
{\bf P}^{N-1} \: = \: \frac{ {\bf C}^N - 0 }{ {\bf C}^{\times} }
\end{displaymath}

We can describe gerbes on projective spaces in a very similar
fashion.  In particular, the ${\bf Z}_k$ gerbe on ${\bf P}^{N-1}$
that generates all of the ${\bf Z}_k$ gerbes is given by the quotient
\begin{displaymath}
\left[ \frac{ {\bf C}^N - 0 }{ {\bf C}^{\times} } \right]
\end{displaymath}
where the ${\bf C}^{\times}$ acts by $k$ times the minimal amount.
The characteristic class of this gerbe is
\begin{displaymath}
-1 \: \in \: {\bf Z}_k \: = \: H^2( {\bf P}^{N-1}, {\bf Z}_k )
\end{displaymath}
essentially because ${\bf C}^N - 0$ with the natural projection to
${\bf P}^{N-1}$ is the total space of ${\cal O}_{ {\bf P}^{N-1} }(-1)$.
Clearly, this corresponds to a gauged linear sigma model with $N$
chiral superfields $x_i$, each of charge $k$ under a gauged $U(1)$.

So, in this example, one describes a gerbe over a projective space,
instead of a projective space, simply by multiplying all the charges
by $k$.

We obtained the description of nonminimal charges from the description
of the quotient, but there is also another way to obtain this description,
and that is directly from the definition of the gerbe.
Described as a quotient stack $[ ( {\bf C}^N - 0 ) / {\bf C}^{\times}]$,
a map into the gerbe is a pair consisting of a principal ${\bf C}^{\times}$
bundle
$L$ over the worldsheet, together with a ${\bf C}^{\times}$-equivariant map
from the total space of $L$ into ${\bf C}^N - 0$, where the ${\bf C}^{\times}$
acts by $k$ times the usual amount.  
Now, mathematically, a pair consisting of a
principal ${\bf C}^{\times}$ bundle $P$ together with a map $P \rightarrow 
{\bf C}$, equivariant with respect to a ${\bf C}^{\times}$ action that rotates
the ${\bf C}$ $k$ times, is equivalent to a pair consisting of a line
bundle $L$ together with a section of $L^{\otimes k}$.
We can see this as follows.
Starting with the pair $(P,\phi : tot(P) \to
{\bf C})$, construct the pair $(L, s \in H^{0}(X,L^{k}))$ by
setting $L = P\times_{\chi} {\bf C}$ and $s(x) =
[(p,\phi(p))]$. Here we have:
\begin{itemize}
\item $\chi : {\bf C}^{\times} \to {\bf
C}^{\times}$ is the character $\alpha \to \alpha^{k}$;
\item $P\times_{\chi}
{\bf C}$ is the associated line bundle. It is given explicitly as
the quotient of $P\times {\bf C}$ by the ${\bf C}^{\times}$
action in which $\alpha \in {\bf C}^{\times}$ acts on $(p,z)$ as
$(p\cdot \alpha^{-1}, \alpha^{-k}z)$.
\item $[(p,\phi(p))] \in L_{x}$ denotes the ${\bf C}^{\times}$
orbit of the point $(p,\phi(p)) \in P_{x}\times {\bf C}$. Note
that $[(p,\phi(p))]$ is independent of the choice of $p \in P_{x}$
since $\phi$ is $\chi$-equivariant.
\end{itemize}
Conversely, given $(L,s)$, take $P$ to be the frame bundle of $L$,
{\it i.e.} $L$ with the zero section removed. The pullback of $L$ to $P$ is
canonically trivialized since the tautological section is a frame. The
pullback of $s$ to $P$ is then a section of the trivial bundle, {\it i.e.} a
map from $P$ to ${\bf C}$. 
It is then straightforward to check that this map is equivariant.
In particular, a set of not-simultaneously-vanishing maps from $tot(P)$ to
${\bf C}$ is equivalent to a set of not-simultaneously-vanishing
Higgs fields.
Thus, the data describing the gerbe as a quotient stack is equivalent
to data describing line bundles with sections of $L^{\otimes k}$
(taking into account D-terms),
which is manifestly the description of the Higgs fields.
Thus, one can recover the nonminimal charge description directly
from the definition of the gerbe.

Before describing how the nonperturbative physics of this two-dimensional
gauge theory differs from that of the ordinary ${\bf C} {\bf P}^{N-1}$
model, let us take a moment to understand why this two-dimensional
gauge theory should correspond to the gerbe.  In the ordinary ${\bf C}
{\bf P}^{N-1}$ model, D-terms restrict the classical Higgs vacua to
a sphere $S^{2N-1}$, so gauging $U(1)$ rotations leaves
\begin{displaymath}
\frac{ S^{2N-1} }{ U(1) } \: = \: {\bf P}^{N-1}
\end{displaymath}
In the present case, we are gauging a $U(1)$ that acts by $k$ rotations
of the $U(1)$ fibers of the principal $U(1)$ bundle over ${\bf P}^{N-1}$
whose total space is the sphere $S^{2N-1}$.  Since the $U(1)$ rotates
$k$ times instead of once, locally this is the same as a trivially-acting
${\bf Z}_k$ orbifold of ${\bf P}^{N-1}$ -- the expectation values
of the Higgs fields do not completely break the gauge symmetry,
but only break it to a noneffectively-acting subgroup.  However, because the
sphere $S^{2N-1}$ is a nontrivial $U(1)$ bundle, the resulting local orbifold
is not the same as the global orbifold $[ {\bf P}^{N-1} / {\bf Z}_k ]$.

We have seen in numerous examples in \cite{nr} that orbifolding by
a noneffectively-acting group does not reproduce the original theory.
To be consistent, one would expect that two-dimensional gauge theories
with nonminimal charges must differ from two-dimensional gauge theories
with minimal charges.

Perturbatively in QFT, multiplying all the charges by a nonzero constant
has absolutely no effect on the theory.  However, 
{\it nonperturatively},
it can have an effect, as the charges determine which bundles each
field couples to, as discussed in section~\ref{expnonmin}.  
For example, in a degree $d$ instanton background,
in the gauged linear sigma model describing merely ${\bf P}^{N-1}$,
the $x_i$ are holomorphic sections of ${\cal O}(d)$,
whereas in the linear sigma model describing the gerbe,
the $x_i$ are holomorphic sections of ${\cal O}(kd)$.

Just as in the ordinary ${\bf C} {\bf P}^{N-1}$ model, where the
axial $U(1)_R$ is broken to ${\bf Z}_{2N}$ by anomalies,
here the same $U(1)_R$ is broken to ${\bf Z}_{2 k N}$.
We can derive this from {\it e.g.} \cite{horivafa}[equ'n (3.4)],
or alternatively, since the low-energy theory can be approximated
by the ${\bf C} {\bf P}^{N-1}$ model with a restriction that degrees of maps
are divided by $k$ (as we shall see explicitly in a few moments),
we can do a Riemann-Roch computation in the worldsheet A model.
Furthermore, following the methods of \cite{edold}, it is natural
to conjecture that the IR limit of this theory develops a mass gap
and has $kN$ vacua, just as the ordinary ${\bf C}{\bf P}^{N-1}$ model 
has a gap and $N$ vacua.  We shall confirm that speculation in
section~\ref{todagpn}, when we derive and analyze the Toda theory
corresponding to this model.

Now that we have begun to see some of the reasons why this theory
is physically distinct from the ordinary ${\bf C} {\bf P}^{N-1}$ model,
let us check the relation between this theory and the mathematics
of gerbes. 

The first check we shall perform involves the Witten index.
In the ordinary ${\bf C} {\bf P}^{N-1}$ model, 
the Witten index $\mbox{Tr }(-)^F$
computes the Euler characteristic of the target, ${\bf P}^{N-1}$.
In the present case, given a mass gap and $kN$ vacua in the IR,
the Witten index of the present theory is $kN$.
Now, as described in more detail in \cite{tonyme}, the relevant notion
of Euler characteristic for stacks is the Euler characteristic of the
associated inertia stack (the ``orbifold Euler characteristic,''
for stacks presented as global quotients by finite groups),
and for the gerbe at hand, the Euler characteristic of the associated
inertia stack is precisely $kN$, in agreement with the Witten index.

Now, let us study linear sigma model
moduli spaces and quantum cohomology, and compare these physical
properties to what one expects mathematically for the gerbe.

The linear sigma model moduli space is obtained by expanding
worldsheet fields in a basis of zero modes.  The coefficients
in the expansion are the homogeneous coordinates on the moduli space,
and are given the same $U(1)$ charges as the original field.
In the present case, for what morally are degree $d$ maps,
we have that $x_i \in \Gamma( {\cal O}(kd))$ for the gauged linear
sigma model with nonminimal charges.
We exclude coefficients such that all the $x_i \equiv 0$,
leaving us with
\begin{displaymath}
{\cal M} \: = \: \left[ \frac{ {\bf C}^{N(kd+1)} - 0 }{ {\bf C}^{\times} }
\right]
\end{displaymath}
where all fields have charge $k$ with respect to the 
${\bf C}^{\times}$.

If the fields all had charge $1$ with respect to the ${\bf C}^{\times}$,
then ${\cal M} = {\bf P}^{N(kd+1) - 1}$.
However, since they have charge $k$,
if we are careful about the quotient,
then we must interpret the moduli `space' ${\cal M}$ as itself
being a stack, and in particular, a ${\bf Z}_k$ gerbe over
${\bf P}^{N(kd+1)-1}$.

Before computing quantum cohomology,
let us compare these physical statements to the mathematics
of maps into gerbes.
First, each map from ${\bf P}^1$ into a ${\bf Z}_k$ gerbe
has a ${\bf Z}_k$'s worth of automorphisms.
If we keep track of those automorphisms,
then our moduli `space' of maps does not have a {\it set} of points,
so much as a {\it category} of points, in which each
point is replaced by a copy of the classifying stack $B {\bf Z}_k$.
This is exactly the local structure of a ${\bf Z}_k$ gerbe.
In other words, on mathematical grounds, because maps from the worldsheet
into the gerbe have automorphisms, we should expect that the
moduli `space' should have the structure of a gerbe,
a ${\bf Z}_k$ gerbe in fact, and that is exactly what we see in
the linear sigma model moduli space.

We can check even more.
Our nonperturbative interpretation of nonminimal charges
ensures that the degree of each map in the moduli space is a 
multiple of $k$.  This physical consequence of nonminimal charges
has a mathematical explanation in terms of gerbes.
Since there is a projection map from the ${\bf Z}_k$ gerbe into
the underlying manifold ${\bf P}^{N-1}$, each map from
the worldsheet ${\bf P}^1$ into the gerbe also defines a map $f$
from ${\bf P}^1$ into ${\bf P}^{N-1}$.
Now, a map from ${\bf P}^1$ into a gerbe ${\cal G}$ is the same
as a map $f$ into ${\bf P}^{N-1}$ together with a trivialization\footnote{
A map from ${\bf P}^1$ into the gerbe ${\cal G}$ over $M$ defines a map
from ${\bf P}^1$ into the fiber product ${\bf P}^1 \times_M {\cal G} = 
f^* {\cal G}$, and such a map is a trivialization.  This is seen perhaps
most easily if we replace gerbes with bundles.  If ${\cal G}$ is a bundle
over a space $X$, and we are given a map $f: {\bf P}^1 \rightarrow X$,
then since there is a canonical map $f^* {\cal G} \rightarrow {\cal G}$,
a map ${\bf P}^1 \rightarrow f^* {\cal G}$ can be composed with that
canonical map to give a map $g: {\bf P}^1 \rightarrow {\cal G}$,
such that $f = \pi \circ g$ for $\pi: {\cal G} \rightarrow X$.
Conversely, given a map $g: {\bf P}^1 \rightarrow {\cal G}$,
if we let $f = \pi \circ g$, then we can define a map ${\bf P}^1 
\rightarrow f^* {\cal G}$ as follows.  Recall
\begin{displaymath}
f^* {\cal G} \: = \: \{ (x, e) \in {\bf P}^1 \times {\cal G} |
f(x) = \pi(e) \}
\end{displaymath}
so given $g: {\bf P}^1 \rightarrow {\cal G}$, we can map
$x \mapsto (x,g(x)) \in f^* {\cal G}$, for $x \in {\bf P}^1$.
The argument for ${\cal G}$ a gerbe instead of a bundle
is virtually identical.
}
of $f^* {\cal G}$.  Equivalently, the natural map
\begin{displaymath}
f^*: \: H^2\left( {\bf P}^{N-1}, {\bf Z}_k \right) \: \longrightarrow \:
H^2\left( {\bf P}^1, {\bf Z}_k \right)
\end{displaymath}
should be identically zero, otherwise the pullback $f^* {\cal G}$
would not have a trivialization.  But 
\begin{displaymath}
H^2\left( {\bf P}^{N-1}, {\bf Z}_k \right) \: = \:
H^2\left( {\bf P}^1, {\bf Z}_k \right) \: = \: {\bf Z}_k
\end{displaymath}
and the natural map $f^*:  {\bf Z}_k \rightarrow {\bf Z}_k$ is multiplication
by the degree of the map.  Hence, for a gerbe on ${\bf P}^{N-1}$ of
characteristic class $n \mbox{ mod } k$,
if we let $d$ denote the degree of the map 
$f: {\bf P}^1 \rightarrow {\bf P}^{N-1}$, then we have that
\begin{displaymath}
d \left( n \mbox{ mod } k \right) \: = \: 0 \mbox{ mod } k
\end{displaymath}
In the present case, $n = -1$, so we have that $d$
must be divisible by $k$.
Conversely, any map from ${\bf P}^1 \rightarrow {\bf P}^{N-1}$
of degree a multiple of $k$,
together with a section of the trivial gerbe on ${\bf P}^1$,
defines a map from the worldsheet ${\bf P}^1$ into the gerbe.

Thus, not only does the fact that the linear sigma model moduli `space'
is a gerbe have a mathematical interpretation in terms of properties
of maps into gerbes, but also the fact that the maps have degree a
multiple of $k$ also naturally agrees with what one expects for gerbes.
Put more simply, the linear sigma model moduli `space' is a compactification
of the moduli space of maps into the gerbe.

Next let us 
outline a physical prediction for the
quantum cohomology of this gerbe, which will also further illustrate
how the physical theory for the gerbe is distinct from the physical
theory for the underlying space.

For the projective space ${\bf P}^{N-1}$, the linear sigma model
moduli space for degree $d$ maps is given by
${\bf P}^{N(d+1)-1}$, so the A model has correlation functions
of the form
$<X^{N(d+1)-1}> = q^d$ (with $X$ corresponding to the generator
of degree two cohomology) defining\footnote{One fast way to think about
the quantum cohomology relations in simple cases is that they
can be used to generate all of the correlators from the classical
correlation functions.  Here, if $X$ generates degree two cohomology,
then since ${\bf P}^{N-1}$ is $(N-1)$-dimensional, the nonzero
classical correlation function
is $<X^{N-1}> = 1$, so we can use the relation $X^N=q$ to derive
the general correlation function above from the classical correlation
function.} a quantum cohomology
relation $X^N = q$.

For the gerbe under consideration, we have seen that the
linear sigma model moduli space is given by a ${\bf Z}_k$ gerbe
over ${\bf P}^{N(kd+1)-1}$, so
from the dimension of the
moduli space we see that in the gerbe theory we have correlators of the form
$<X^{N(kd+1) - 1}> = q^d$ (with $X$ a pullback to the gerbe of
degree two cohomology), defining a quantum cohomology relation
$X^{kN} = q$.   
Since the quantum cohomology ring of this gauged linear sigma model
with nonminimal charges is not isomorphic to the quantum cohomology
ring of the underlying space ${\bf P}^{N-1}$, or more simply,
the correlation functions are different, we see that the
physical theory corresponding to the gerbe
really is different from the theory with minimal charges.

There is another way to derive this quantum cohomology relation in the
gerbe theory, using methods of \cite{ronendave2}. 
By using one-loop effective actions, the authors of \cite{ronendave2} derived
a general expression for quantum cohomology for toric targets from
the gauged linear sigma model.  Applying \cite{ronendave2}[equ'n (3.44)]
to the present example, we find 
\begin{displaymath}
\prod_1^N \left( k \sigma \right)^k \: = \: \exp(2 \pi i \tau)
\end{displaymath}
or, after appropriate redefinitions, we have that
\begin{displaymath}
\sigma^{kN} \: = \: q
\end{displaymath}
the same quantum cohomology relation we derived above.

\subsection{Alternate presentation of the first example}  \label{alt1}

In this section, let us discuss an alternate presentation of the
gerbe discussed in the previous section, as a check that the
physics captured mathematically by the stack
is presentation-independent.

Consider a gauged linear sigma model with $N$ chiral superfields
$x_i$, and another chiral superfield $z$, with two gauged $U(1)$'s,
under which the chiral superfields have charges
\begin{center}
\begin{tabular}{c|cc}
 $\,$ & $x_i$ & $z$ \\ \hline
$\lambda$ & $1$ & $-1$ \\
$\mu$ & $0$ & $k$
\end{tabular}
\end{center}
Associated to each $U(1)$
is a D-term:
\begin{eqnarray*}
\sum_i |x_i|^2 \: - \: |z|^2 & = & r_{\lambda} \\
k |z|^2 & = & r_{\mu}
\end{eqnarray*}
We shall assume that $r_{\mu} \gg 0$, so that $z \neq 0$.
In this case, were we not gauging the second $U(1)$,
the resulting toric variety would be the total space of the
principal $U(1)$ bundle over ${\bf P}^{N-1}$ of degree $-1$.
By gauging the second $U(1)$, we are gauging rotations of the
fiber of that principal $U(1)$ bundle.
If $k=1$, the resulting toric variety is, mathematically,
${\bf P}^{N-1}$.  If $k > 1$, then mathematically we have a ${\bf Z}_k$ gerbe
on ${\bf P}^{N-1}$, the same gerbe as that discussed in the
last section.

Let us study the physics of this gauged linear sigma model,
and check that if $k>1$, not only is the nonperturbative
physics distinct from
that of the ${\bf P}^{N-1}$ model, but also that the A model
twist of this gauged linear sigma model has isomorphic properties
to the model discussed in the last section (so that the physics is
presentation-independent, essentially).

In fact, before doing detailed calculations,
we can immediately perform a quick test\footnote{We would like to thank
K.~Hori for suggesting this test to us.} that this
really is an alternate presentation of the first example.
Note that by redefining ${\bf C}^{\times}$ actions we can map 
this model to one described by homogeneous coordinates with weights
as follows:
\begin{center}
\begin{tabular}{c|cc}
 & $x_i$ & $z$ \\ \hline
$\lambda^k \mu$ & $k$ & $0$ \\
$\mu$ & $0$ & $k$ \\
\end{tabular}
\end{center}
This is {\it almost} the same as the model we started with,
except for the extra homogeneous coordinate.  In fact, this model
describes the previous presentation of the gerbe, tensored with a single
homogeneous coordinate modded out by a ${\bf C}^{\times}$ that acts
with weight $k$ -- which is a presentation of the trivial gerbe $B {\bf Z}_k
 = [ \mbox{point}/{\bf Z}_k ]$.  Now, our map between ${\bf C}^{\times}$
actions has nontrivial kernel; schematically, we have a short exact
sequence
\begin{displaymath}
0 \: \longrightarrow \: {\bf Z}_k \: \longrightarrow \:
\left( \lambda, \mu \right) \: \longrightarrow \:
\left( \lambda^k \mu, \mu \right) \: \longrightarrow \: 0
\end{displaymath}
so, we can relate the theories defined by the two sets of
${\bf C}^{\times}$ actions as
\begin{displaymath}
\left( \lambda^k \mu, \mu \right) \: = \:
\left[ \frac{ \left( \lambda, \mu \right) }{{\bf Z}_k} \right]
\end{displaymath}
The ${\bf Z}_k$ quotient on the right side of the equation above
acts trivially, globally, and as discussed in \cite{tonyme},
the corresponding physical theory of such a global trivial quotient
is a tensor product
\begin{displaymath}
\left( \lambda^k \mu, \mu \right) \: = \:
\left( \lambda, \mu \right) \otimes [ \mbox{point}/{\bf Z}_k ]
\end{displaymath}
In particular, note this is precisely consistent with our underlying
conjecture that stacks classify universality classes of gauged sigma models.
The $(\lambda^k \mu, \mu)$ theory is, explicitly, a tensor product of the
theory defined by the first presentation of the ${\bf Z}_k$ gerbe
over ${\bf P}^{N-1}$, with a theory defined by a presentation of the trivial
gerbe $B {\bf Z}_k = [ \mbox{point}/{\bf Z}_k]$.  
Our analysis of ${\bf C}^{\times}$ actions reveals that the 
$(\lambda^k \mu, \mu)$ theory is also a tensor product of the theory defined
by the $(\lambda,\mu)$ presentation of the ${\bf Z}_k$ gerbe over
${\bf P}^{N-1}$ with a presentation of the trivial gerbe $B {\bf Z}_k$.
These two descriptions of the $(\lambda^k, \mu)$ theory have matching
universality classes, so long as stacks really do classify universality
classes.

Now that we have analyzed ${\bf C}^{\times}$ actions, let us turn to
other tests of our conjecture that stacks classify universality classes
of gauged sigma models.

The linear sigma model moduli space is now defined by two
integers $d_1$, $d_2$, associated to the gauged $U(1)$'s $\lambda$,
$\mu$, respectively.  
The zero modes of the chiral superfields are given by
\begin{eqnarray*}
x_i & \in & \Gamma\left( {\cal O}(d_1 ) \right) \\
z & \in & \Gamma\left( {\cal O}(-d_1 + kd_2) \right)
\end{eqnarray*}
Let us assume for simplicity that $d_1 \geq 0$ and $-d_1 + kd_2 \geq 0$.
Then, following the same analysis as before,
the linear sigma model moduli `space' looks like a ${\bf Z}_k$ gerbe over
${\bf P}^{N(d_1 +1)-1 } \times {\bf P}^{-d_1 + kd_2 }$,
where the homogeneous coordinates $x_{ia}$ and $z_b$ (from the zero
modes of the $x_i$ and $z$) have weights under two $U(1)$'s as
\begin{center}
\begin{tabular}{c|cc}
 $\,$ & $x_{ia}$ & $z_b$ \\ \hline
$\lambda$ & $1$ & $-1$ \\
$\mu$ & $0$ & $k$
\end{tabular}
\end{center}
From the dimensions of the moduli spaces,
correlators in an A model twist of a lower energy theory look like
they should have two generators, $X$, $Y$, with correlation functions
\begin{displaymath}
< X^{N(d_1+1)-1}Y^{-d_1+kd_2}> \: = \: q_1^{d_1} q_2^{d_2}
\end{displaymath}
which given the classical correlation function
$<X^{N-1}>=1$ define quantum cohomology relations
\begin{eqnarray*}
X^N Y^{-1} & = & q_1 \\
Y^k & = & q_2 
\end{eqnarray*}
Note, however, that the quantum cohomology ring with two generators
$X$, $Y$ and relations as above is isomorphic to the ring with
one generator $X$ and relation $X^{kN} = q$ (for $q=q_2 q_1^k$).
In other words, the quantum cohomology ring of this theory
is isomorphic to the quantum cohomology ring of the 
theory of section~\ref{basicpngerbe}.
This means that the A model correlators
are isomorphic between these two theories, exactly as one would
expect if they lie in the same universality class.

Let us check this result by using the methods of \cite{ronendave2}.
Applying \cite{ronendave2}[equ'n (3.44)] to the present case we have
\begin{eqnarray*}
\left( \prod_1^N \sigma_1 \right) \left( - \sigma_1 + k \sigma_2 \right)^{-1}
& = & \exp( 2 \pi i \tau_1 ) \\
\left( - \sigma_1 + k \sigma_2 \right)^k & = & \exp(2 \pi i \tau_2 )
\end{eqnarray*}
By making the identifications
\begin{eqnarray*}
X & \sim & \sigma_1 \\
Y & \sim & - \sigma_1 + k \sigma_2 
\end{eqnarray*}
we recover the quantum cohomology relations derived previously.

\subsection{More gerbes on projective spaces and alternative presentations}
\label{moregerbes}

Let us begin by describing a ${\bf Z}_k$ gerbe on ${\bf P}^{N-1}$
with characteristic class in $H^2(S^2, {\bf Z}_k)$ given by
$-n \mbox{ mod } k$, {\it i.e.} the $G^k_{-n} {\bf P}^{N-1}$ model.

Mathematically, we can describe this gerbe as a ${\bf C}^{\times}$
quotient of the total space of a principal ${\bf C}^{\times}$ bundle
over ${\bf P}^{N-1}$, with the property that 
$c_1 \mbox{ mod } k = -n \mbox{ mod } k$,
in which the ${\bf C}^{\times}$ rotates the fibers $n$ times.

To describe that quotient, consider a linear sigma model with
$N$ chiral superfields $x_i$ plus $z$, and two ${\bf C}^{\times}$
actions, with weights as shown:
\begin{center}
\begin{tabular}{c|cc}
 & $x_i$ & $z$ \\ \hline
$\lambda$ & $1$ & $-n$ \\
$\mu$ & $0$ & $k$ \\
\end{tabular}
\end{center}
We shall assume that $n$ is positive.
The D-terms are
\begin{eqnarray*}
\sum_i |x_i|^2 \:  - \: n |z|^2 & = & r_{\lambda} \\
k |z|^2 & = & r_{\mu}
\end{eqnarray*}
We shall assume that $r_{\mu} \gg 0$ and $r_{\lambda} \gg 0$.
If $r_{\mu} \neq 0$, then $z \neq 0$, 
and the remaining D-term disallows $x_i=0$ for all $i$.
Thus, we are quotienting 
\begin{displaymath}
\left( {\bf C}^N - 0 \right) \times {\bf C}^{\times}
\end{displaymath}
by a pair of ${\bf C}^{\times}$ actions.

Mathematically, this gauged linear sigma model should describe
a gerbe over
${\bf P}^{N-1}$ classified by $-n \mbox{ mod } k$, and we shall
check that physical properties of this gauged linear sigma model
are reflected in mathematical properties of the gerbe.

Since shifting $n \mapsto n + k$ changes the presentation but leaves
the gerbe invariant, the statement that stacks classify universality
classes implies that the pertinent physics should be invariant
under $n \mapsto n + k$, and we shall check that statement.

Let us consider curve counting in this gerbe, following the same
procedure as in the last few examples.
First, note that a single integer no longer suffices to specify the
degree of the maps, we must specify two integers $d_1$, $d_2$.
Then, expanding fields in zero modes to get the linear sigma model
moduli space, we have
\begin{eqnarray*}
x_i & \in & H^0\left( {\bf P}^1, {\cal O}(d_1) \right) \: = \:
{\bf C}^{d_1 + 1} \\
z & \in & H^0\left( {\bf P}^1, {\cal O}(-n d_1 + k d_2) \right)
\: = \: {\bf C}^{-n d_1 + k d_2 + 1}
\end{eqnarray*}
(For simplicity, we shall assume that $d_1 \geq 0$ and
$-kd_1 + nd_2 \geq 0$.)
The resulting moduli space ${\cal M}$ 
can be described in GLSM language as $N(d_1+1)$ homogeneous
coordinates $x_{ia}$, $-nd_1 + kd_2 + 1$ homogeneous coordinates $z_b$,
with a pair of ${\bf C}^{\times}$ actions under which the homogeneous
coordinates have weights
\begin{center}
\begin{tabular}{c|cc} \\
 & $x_{ia}$ & $z_b$ \\ \hline
$\lambda'$ & $1$ & $-n$ \\
$\mu'$ & $0$ & $k$ \\
\end{tabular}
\end{center}
The resulting linear sigma model moduli `space'
is a ${\bf Z}_k$-gerbe over a 
projectivization
of the vector bundle ${\cal O}(-n)^{ \oplus -nd_1 + k d_2 + 1}$
over ${\bf P}^{N(d_1+1)-1}$, or, more simply,
a ${\bf Z}_k$ gerbe over ${\bf P}^{-n d_1 + k d_2 } 
\times {\bf P}^{N(d_1+1)-1}$.

As before, given the moduli space, in a simple
example of this
form we can now compute quantum cohomology in the A model twist
of a lower-energy theory.
Correlators have two generators $X$, $Y$, and from the dimension of the
moduli space we can read off the correlation functions
\begin{equation}    \label{gnpcorr}
< X^{N(d_1+1)-1}Y^{-nd_1 + kd_2 }> \: = \: q_1^{d_1} q_2^{d_2} 
\end{equation}
From these correlation functions it is clear that with respect to
the generators $X$, $Y$, the quantum cohomology ring is defined
by the relations
\begin{eqnarray*}
X^N Y^{-n} & = & q_1 \\
Y^k & = & q_2
\end{eqnarray*}
Note that we immediately recover the previous example as a special case.

Note that if we shift $n \mapsto n + k$, which leaves the
gerbe invariant, the resulting moduli space is equivalent to the
moduli space with 
$d'_1 = d_1$ and $d'_2 = -d_1 + d_2$.
Since physically we sum over all degrees, A model physics
can only depend upon $n \mbox{ mod } k$.  Similarly,
under $n \mapsto n + k$, the quantum cohomology ring is
invariant -- the new relations are equivalent to the old ones,
with $q_1' = q_1/q_2$.
Thus, A model results only depend upon $n$ and $k$ in the combination
$n \mbox{ mod } k$, consistent with the assumption that
universality classes of gauged sigma models are classified by
stacks, not by presentations thereof.

Let us check this quantum cohomology calculation by
using the methods of \cite{ronendave2}.
Applying \cite{ronendave2}[equ'n (3.44)] to the present case
gives us
\begin{eqnarray*}
\left( \prod_1^N \sigma_1 \right) \left( -n \sigma_1 \: + \: k \sigma_2 \right)^{-n} & = &
\exp(2 \pi i t_1 ) \\
\left( -n \sigma_1 \: + \: k \sigma_2 \right)^k & = & 
\exp(2 \pi i t_2 )
\end{eqnarray*}
If we make the identifications
\begin{eqnarray*}
X & \sim & \sigma_1 \\
Y & \sim & -n \sigma_1 + k \sigma_2
\end{eqnarray*}
then we recover the previous expression for quantum cohomology.

The reader should note that the quantum cohomology relations we have
derived give a product structure on the cohomology of the inertia
stacks associated to these gerbes, which all look like $k$ disjoint
copies of the gerbe.  In computing quantum cohomology, starting only
with what should be untwisted sector fields, we have implicitly recovered
twist fields as well.  Also note this provides a check of the claim
in \cite{tonyme} that the massless spectrum of a string on a stack
should be counted by cohomology of the inertia stack.

In passing, note that when $n=0$, {\it i.e.} when the gerbe is trivial, 
the quantum cohomology ring above
is a product of the quantum cohomology of the ordinary ${\bf P}^{N-1}$
model and a twist field, agreeing with general results on factorizability
of physical theories associated to trivial gerbes presented in
\cite{nr}.

\section{More general toric stacks}    \label{moreexs}

For the most part we concentrate in this paper on the special
case of gauged linear sigma models for
${\bf Z}_k$ gerbes over projective spaces,
as in our opinion those give very clear illustrations of the
pertinent physics.  Gauged linear sigma models for more general
toric stacks are very similar, in that they typically look like
gauged linear sigma models on toric varieties, except that the
charges with respect to some of the $U(1)$'s are nonminimal.
The only exceptions are those toric stacks whose GLSM description
is identical to that traditionally associated to certain special
toric varieties, as we shall discuss in subsection~\ref{wps}.

\subsection{Stacks over Hirzebruch surfaces}

Just to give some easy examples of more general toric stacks and their
corresponding gauged linear sigma models, let us consider Hirzebruch
surfaces.  Recall that these surfaces, traditionally denoted ${\bf F}_n$
and indexed by integers, correspond to total spaces of ${\bf P}^1$-bundles
over ${\bf P}^1$'s.  They can be described with gauged linear sigma models
very simply as follows.  Let $s$, $t$, $u$, $v$ be homogeneous coordinates;
then the gauged linear sigma model has two $U(1)$'s, 
call them $\lambda$, $\mu$, under which the 
homogeneous coordinates have charges as follows:
\begin{center}
\begin{tabular}{c|cccc}
 & $s$ & $t$ & $u$ & $v$ \\ \hline
$\lambda$ & $1$ & $1$ & $n$ & $0$ \\
$\mu$ & $0$ & $0$ & $1$ & $1$\\
\end{tabular}
\end{center}
The homogeneous coordinates $s$, $t$ act as homogeneous coordinates
on the base of the ${\bf P}^1$ bundle, and the coordinates $u$, $v$ act
as homogeneous coordinates on the fiber of the ${\bf P}^1$ bundle.

Suppose you wanted to describe a ${\bf Z}_k$ gerbe over the base of the
${\bf P}^1$ bundle.  Then, a gauged linear sigma model for such a toric stack
would have homogeneous coordinates $s$, $t$, $u$, $v$, with charges
\begin{center}
\begin{tabular}{c|cccc}
 & $s$ & $t$ & $u$ & $v$ \\ \hline
$\lambda$ & $k$ & $k$ & $kn$ & $0$ \\
$\mu$ & $0$ & $0$ & $1$ & $1$\\
\end{tabular}
\end{center}
Similarly, a ${\bf Z}_k$ gerbe over the fiber of the ${\bf P}^1$ bundle
could be described as
\begin{center}
\begin{tabular}{c|cccc}
 & $s$ & $t$ & $u$ & $v$ \\ \hline
$\lambda$ & $1$ & $1$ & $n$ & $0$ \\
$\mu$ & $0$ & $0$ & $k$ & $k$\\
\end{tabular}
\end{center}
and so forth.

\subsection{Analogue of flops}

In gauged linear sigma models studied in the past,
varying the Fayet-Iliopoulos parameters through different geometric
phases has the effect of realizing birational transformations.
The same statement is true of gauged linear sigma models describing 
toric stacks, for the basic reason that varying the Fayet-Iliopoulos parameters
only changes the exceptional set of the quotient, so on a Zariski-open
subset the resulting stacks will be isomorphic.

Let us illustrate this with a gerby analogue of a classic example
from \cite{phases} -- flopping between the two small resolutions of
a conifold singularity.

Consider the linear sigma model with five chiral superfields
$a$, $b$, $c$, $d$, $e$, and ${\bf C}^{\times}$ actions as below:
\begin{center}
\begin{tabular}{c|ccccc}
 & $a$ & $b$ & $c$ & $d$ & $e$ \\ \hline
$\lambda$ & $1$ & $1$ & $-1$ & $-1$ & $k$ \\
$\mu$ & $0$ & $0$ & $0$ & $0$ & $n$ \\
\end{tabular}
\end{center}
If we omitted $e$ and $\mu$, then we would recover the small
resolutions of the affine conifold.
The corresponding D terms are
\begin{eqnarray*}
|a|^2 \: + \: |b|^2 \: - \: |c|^2 - |d|^2 + k |e|^2 & = & r_{\lambda} \\
n |e|^2 & = & r_{\mu}
\end{eqnarray*}
We shall assume that $r_{\mu} \gg 0$, so that $e \neq 0$.
Integrating out $e$ leaves us with a single D-term given by
\begin{displaymath}
|a|^2 \: + \: |b|^2 \: - \: |c|^2 - |d|^2
\: = \: r_{\lambda} \: - \: (k/n) r_{\mu}
\end{displaymath}

When $r_{\lambda} - (k/n) r_{\mu} \ll 0$, the D-term dictates an exceptional
set $\{ c=d=0 \}$, so that we are left with the quotient
\begin{displaymath}
\frac{ {\bf C}^2 \times \left( {\bf C}^2 - 0 \right) \times {\bf C}^{\times} }{
{\bf C}^{\times} \times {\bf C}^{\times} }
\end{displaymath}
which appears to be a ${\bf Z}_n$ gerbe over one small resolution of
the conifold.

When $r_{\lambda} - (k/n) r_{\mu} \gg 0$, the D-term gives the exceptional
set $\{ a=b=0\}$.
We are left with the quotient
\begin{displaymath}
\frac{ \left( {\bf C}^2 - 0 \right) \times {\bf C}^2 \times {\bf C}^{\times} }{ {\bf C}^{\times} \times
{\bf C}^{\times} }
\end{displaymath}
describing a gerbe over the other small resolution.

\subsection{Weighted projective stacks}  \label{wps}

We have so far discussed toric stacks that can be described by
gauged linear sigma models with nonminimal charges.  However,
there are some toric stacks whose corresponding gauged linear sigma models
look identical to that traditionally associated to the underlying space.

The easiest examples of such toric stacks are the weighted projective
stacks.  These are defined in exactly the same way as weighted projective
spaces, except that one takes a stacky quotient instead of an ordinary
quotient.  For example, a weighted projective space
$W {\bf P}^n_{k_1,k_2, \cdots}$ is defined as the quotient
\begin{displaymath}
\frac{ {\bf C}^{n+1} - \{ 0 \} }{ {\bf C}^{\times} }
\end{displaymath}
where the ${\bf C}^{\times}$ acts on the homogeneous coordinates
with weights $(k_1, k_2, \cdots)$.
The corresponding weighted projective stack is defined in almost
exactly the same way, as the stack quotient
\begin{displaymath}
\left[ 
\frac{ {\bf C}^{n+1} - \{ 0 \} }{ {\bf C}^{\times} }
\right]
\end{displaymath}
where the ${\bf C}^{\times}$ acts on the homogeneous coordinates
with weights $(k_1, k_2, \cdots)$, exactly the same as for the
weighted projective space.
The local quotient singularities of the weighted projective space
are replaced by stacky structures in the weighted projective stack.

As a trivial example, recall one of the presentations of the
$G^k_{-1} {\bf P}^{N-1}$ gerbe is as a weighted projective
stack, which could be called $W{\bf P}^{N-1}_{[k,k,\cdots, k]}$.

Traditionally, gauged linear sigma models with a single $U(1)$ and
minimal charges $(k_1, k_2, \cdots)$ are believed to be associated
to weighted projective spaces.  However, the present work makes us suspect
that, away from coupling extremes, such gauged linear sigma models might
be more sensibly associated to weighted projective stacks, instead of spaces.
One way to test this conjecture might be to examine quantum cohomology
computations \`a la \cite{ronendave2}, and compare them to
results obtained by other methods for ordinary
weighted projective spaces (which must first be resolved before
the quantum cohomology can be defined).  
We shall not speculate further on this subtle distinction of interpretations
in this paper.

\subsection{Quantum cohomology for toric stacks}

There is now an easy prediction for part of the quantum cohomology
of a toric stack.  In \cite{ronendave2}, predictions for quantum
cohomology of toric varieties were derived by computing one-loop corrections
to effective actions in gauged linear simga models.  However, the authors
of \cite{ronendave2} were aware \cite{ronenpriv} of the physical distinction 
between gauged linear sigma models with minimal and nonminimal charges,
and although they did not understand the mathematical interpretation,
were careful to write down results also valid for gauged linear sigma models
with nonminimal charges.  Thus, on the face of it,
the quantum cohomology calculations in \cite{ronendave2} should also
apply equally well to toric stacks, not just toric varieties.
Indeed, we have checked that statement in previous sections by
comparing several calculations of quantum cohomology for ${\bf Z}_k$
gerbes on projective spaces, some of those calculations coming from
\cite{ronendave2}.

Now, we should qualify this statement slightly.
We have argued in \cite{tonyme} that massless spectra should be
computed by associated inertia stacks.  Gauged linear sigma models, however,
only seem to have direct UV access to the untwisted part of the 
associated inertia stack.  Thus, one expects that
the quantum cohomology predictions of
\cite{ronendave2} would not be predictions for the full quantum cohomology
ring of a toric stack, but only part of it.  Curiously, despite that
expectation, the quantum cohomology rings we derived previously for
${\bf Z}_k$ gerbes over projective spaces were product structures on the
cohomology of the entire inertia stack, not just one sector.
Although the gauged linear sigma models only have direct UV access
to part of the inertia stack, the quantum cohomology relations that one
derives at least sometimes seem to know about all of the inertia stack.
It is not clear whether that will be the case for all toric varieties,
however.

In \cite{agv}, a proposal was made for quantum cohomology rings of
stacks.  Their proposal was based on purely mathematical extrapolations
of existing quantum cohomology calculations, and made the assumption
(not checked until \cite{tonyme}) that the right notion of massless spectrum
should be given by inertia stacks.  It would be interesting to compare their
proposal for quantum cohomology of stacks to the physics results outlined
in this paper.

\section{Calabi-Yau gerbes}   \label{cygerbes}

In \cite{tonyme} the notion of ``Calabi-Yau'' for stacks is
discussed extensively, and it is argued that the `correct' notion
is the naive one:  a stack should be said to be Calabi-Yau if
its canonical bundle is trivial.
For Calabi-Yau gerbes, this constraint turns out to be rather
trivial, as the canonical bundle of the gerbe is just a pullback of
the canonical bundle of the underlying variety.
Thus, a gerbe is Calabi-Yau if and only if the underlying variety
is Calabi-Yau.

In this section we shall discuss some Calabi-Yau gerbes.

\subsection{Example:  threefold in $G_{-1}^k {\bf P}^4$ and Landau-Ginzburg
orbifolds}   \label{lgorb}

One easy way to construct some examples of gauged-linear-sigma-model
presentations of gerbes over Calabi-Yau manifolds is to start with
a gauged linear sigma model describing the Calabi-Yau,
and multiply the $U(1)$ charges of all fields by a constant,
much as in the example discussed in section~\ref{basicpngerbe}.
As already discussed, although perturbatively multiplying the charges
of all the fields by a constant has no effect,
{\it nonperturbatively} the resulting theory is different,
so the full physical theory on a gerbe (for nonzero gauge coupling)
is distinct from the sigma model on the underlying manifold.

For example, consider the quintic hypersurface in ${\bf P}^4$.
As in \cite{phases}, this is described by chiral superfields
$x_i$ of charge 1 under a $U(1)$, together with a chiral superfield
$p$ of charge $-5$, and a superpotential of the form
\begin{displaymath}
W \: = \: p G(x_i)
\end{displaymath}
where $G$ is the homogeneous polynomial of degree 5 defining the
hypersurface.  If we multiply the charges of all fields by $k$,
so that the $x_i$ have charge $k$ and $p$ has charge $-5k$,
then the superpotential remains gauge-invariant, and the D-terms
have the same form as before.

Clearly this process can be repeated for any Calabi-Yau,
reflecting the fact discussed in \cite{tonyme}
that all (Deligne-Mumford) gerbes over Calabi-Yau manifolds
are themselves Calabi-Yau.

The phases of this gauged linear sigma model have the same
general form as for the ordinary quintic.  For large positive $r$,
we recover a Calabi-Yau hypersurface in $G^k_{-1} {\bf P}^4$,
which is a gerbe over the quintic.

For large negative $r$, we recover a Landau-Ginzburg orbifold phase.
In this phase, $p \neq 0$, and its vacuum expectation value breaks
the $U(1)$ to ${\bf Z}_{5k}$.  If we let $\xi$ denote a generator of the
$(5k)$th roots of unity, then the residual gauge group action on the
fields $x_i$ is generated by
\begin{displaymath}
x_i \: \mapsto \: \xi^k x_i
\end{displaymath}
for each $i$, since each $x_i$ has charge $k$.  As a result, although the
orbifold group is ${\bf Z}_{5k}$, only ${\bf Z}_{5k}/{\bf Z}_k = 
{\bf Z}_5$ acts effectively.  Thus, the Landau-Ginzburg orbifold
corresponding to this Calabi-Yau gerbe is a noneffective orbifold.

Orbifolds by noneffectively-acting groups are discussed
extensively in \cite{tonyme}.  Although only the ${\bf Z}_5$ acts
effectively, the physical theory is significantly different from
the ${\bf Z}_5$ orbifold.  In particular, we can now calculate
the massless spectrum, and essentially because ${\bf Z}_{5k}$ is
abelian, we can immediately read off that the massless spectrum
should be given by $k$ copies of the massless spectrum of the
${\bf Z}_5$ Landau-Ginzburg orbifold that corresponds to an
ordinary quintic -- for each twisted sector of the 
${\bf Z}_5$ Landau-Ginzburg orbifold, we get $k$ copies
in the present noneffective orbifold.

It is clear, in fact, that one will obtain results of a similar
character for Landau-Ginzburg orbifolds corresponding to
hypersurfaces in any weighted projective space.

Now, we can use this calculation to provide an important check
of the closed string massless spectrum conjecture presented in
\cite{tonyme}.  There, we conjectured that the closed string massless
spectrum of the IR fixed point of a gauged sigma model was given
by the cohomology of the associated inertia stack.  This agrees with
standard results for orbifolds by finite effectively-acting groups,
and we argue extensively in \cite{tonyme} that this is also the
correct result for orbifolds by finite non-effectively acting groups.
However, for stacks that cannot be presented as quotients by finite groups,
it is not currently possible to directly calculate the massless spectrum of the
IR fixed point, as all presentations are as massive UV theories.
Landau-Ginzburg orbifolds provide a workaround for this technical issue.
Although the Landau-Ginzburg orbifold spectrum need not be precisely
the same as that of the large-radius theory, in typical cases it is
closely related.

Now, for this example of a ${\bf Z}_k$ gerbe over the quintic, 
the associated inertia stack
is $k$ disjoint copies of the gerbe, and so the cohomology of the
inertia stack is given by $k$ copies of the cohomology of the quintic.
We computed above that the massless spectrum of the Landau-Ginzburg
orbifold associated to our Calabi-Yau gerbe is given by $k$ copies
of the massless spectrum of the Landau-Ginzburg ${\bf Z}_5$ orbifold 
associated to the ordinary quintic, in perfect agreement with 
the conjecture that massless spectra should be computed by the inertia stack.
The same result follows for other easy examples of Calabi-Yau hypersurfaces.
Thus, by computing spectra at Landau-Ginzburg points, we have obtained
very strong evidence for the conjecture that massless spectra of 
strings on stacks should be counted by the inertia stack, in the case
of stacks that cannot be presented as global quotients by finite groups.

\subsection{Example:  Calabi-Yau in $G_{-n}^k {\bf P}^{N-1}$ and GLSM phases}

In the previous section we discussed Calabi-Yau stacks built physically
from gauged linear sigma models with nonminimal charges.
As discussed earlier, we can also describe gerbes and stacks as
{\it e.g.} $U(1)$ quotients of total spaces of principal $U(1)$
bundles.  In this section, we shall outline the relevant physics of
such descriptions.

Recall that banded ${\bf Z}_k$ gerbes over ${\bf P}^{N-1}$,
{\it i.e.} the $G^k_{-n} {\bf P}^{N-1}$ model, can be described
as a gauged linear sigma model with $N$ chiral superfields $x_i$ and
one chiral superfield $z$ with charges under a pair of $U(1)$'s as
\begin{center}
\begin{tabular}{c|cc}
 & $x_i$ & $z$ \\ \hline
$\lambda$ & $1$ & $-n$ \\
$\mu$ & $0$ & $k$ \\
\end{tabular}
\end{center}
with D-terms constraining the $z$ to be nonzero.
A Calabi-Yau hypersurface in such an object is $\{ z G(x_i) = 0\}$,
where $G(x_i)$ is of degree $N$ in the $x_i$.
Because of the D-term constraint, there is no
new $\{ z=0 \}$ branch, the solutions of $z G(x_i) = 0$
are the same as the solutions of $G(x_i) = 0$.
Clearly, for any Calabi-Yau presented as a hypersurface in a projective
space, we can trivially construct a Calabi-Yau gerbe,
in agreement with general observations \cite{tonyme} that
a Calabi-Yau gerbe is merely a gerbe over a Calabi-Yau variety.

There is an obvious way to build a full gauged linear sigma model with
a superpotential realizing the Calabi-Yau described above.
Following the usual procedure, add a chiral superfield $p$
with charges $(n-N,-k)$ under $(\lambda,\mu)$.
The superpotential of the theory is given by $pzG(x_i)$,
and we have D-terms given by
\begin{eqnarray*}
\sum_i |x_i|^2 \: - \: n |y|^2 \: + \: (n-N)|p|^2 & = & r_{\lambda} \\
k |y|^2 \: - \: k |p|^2 & = & r_{\mu}
\end{eqnarray*}

The analysis of the GLSM phases is straightforward.
For $n \leq N$, we have a geometric phase, where the
Fayet-Iliopoulos terms are given by $r_{\lambda} \gg 0$,
$r_{\mu} \gg 0$.  In that regime, assuming a smooth hypersurface,
$y \neq 0$, $p=0$, and not all the $x_i$ vanish.

However, we do not seem to ever have an ordinary Landau-Ginzburg
phase in these models, merely hybrid Landau-Ginzburg phases at best.
For example, when $r_{\lambda} \ll 0$ and $r_{\mu} \gg 0$,
we have $y \neq 0$, and either $p = 0$ (and some
$x_i$ nonzero) or all of the $x_i = 0$ (and $p \neq 0$).
The second branch would correspond to a Landau-Ginzburg phase,
but because of the first branch, this is not what one would
ordinarily call an honest Landau-Ginzburg point.

Also note that if $n > N$, then we do not seem to have a purely
geometric phase any longer.  In the regime $r_{\lambda} \gg 0$
we are merely guaranteed that not all of the $\{ x_i, p \}$ vanish.
Thus, in this phase, we have a new branch.

\section{Mirrors to stacks}    \label{mirrors}

In this section we will discuss how gauged-linear-sigma-model-based
methods of \cite{ronendave,horivafa} for computing mirrors
can be applied to the gauged linear sigma model descriptions of
gerbes discussed in this paper.
We will begin with the A model on gerbes on projective spaces
and derive corresponding Toda theories,
and then discuss the mirror of strings on gerbes on quintic threefolds.

\subsection{Toda theories corresponding to gerbes on projective spaces}
\label{todagpn}

For simplicity, we shall begin with gerbes on ${\bf P}^1$,
and shall derive Toda theories from several presentations.
Then, we shall briefly outline gerbes on more general projective
spaces.

\subsubsection{Toda theory for $G_{-1}^k {\bf P}^{N-1}$}

Recall that the $G^k_{-1} {\bf P}^{N-1}$ model is described by
$N$ chiral superfields $x_i$, each of charge $k$ with respect to
a gauged $U(1)$.  In other words, it can be presented
identically to the ${\bf C} {\bf P}^{N-1}$ model, except that the
gauge charges are minimal.

Following the prescription of \cite{ronendave,horivafa}, the
`mirror' theory should be described by $N$ neutral chiral superfields
$Y_i$, with one gauge multiplet $\Sigma$, and an effective superpotential
\begin{displaymath}
\widetilde{W} \: = \: \Sigma\left( k Y_1 \: + \: \cdots \: + \:
k Y_N \right) \: + \: \sum_{i=1}^N \exp(-Y_i)
\end{displaymath}
where we are being sloppy about FI terms and factors, which will
not play an essential role in our discussion.
Integrating out the $Y_i$ yields an effective superpotential for
$\Sigma$ identical to that obtained by one-loop calculations in
\cite{ronendave2}.  If instead we integrate out $\Sigma$,
then in principle we are left with an effective Landau-Ginzburg theory
whose B model correlators should match the A model correlation functions
of the $G^k_{-1} {\bf P}^{N-1}$ model.

Integrating out $\Sigma$ is slightly subtle.  
It gives the constraint
\begin{displaymath}
k \left( Y_1 \: + \: \cdots \: + \: Y_N \right) \: = \: 0
\end{displaymath}
Because the $Y_i$ are periodic, it is not quite right to divide out the $k$.
Rather, this constraint says that $\exp(-Y_N)$ only matches
\begin{displaymath}
\exp(Y_1) \exp(Y_2) \cdots \exp(Y_{N-1})
\end{displaymath}
up to a $k$th root of unity, call it $\Upsilon$.
Thus, integrating out $\Sigma$ gives us an effective superpotential
\begin{displaymath}
\widetilde{W} \: = \: \exp(-Y_1) \: + \: \cdots \:
\exp(-Y_{N-1}) \: + \:
\Upsilon \prod_{i=1}^{N-1} \exp(Y_i)
\end{displaymath}
We will derive this same result from an alternative presentation
of the gerbe in the next section.

\subsubsection{Toda theory for $G_{-n}^k {\bf P}^{N-1}$}

Recall that the $G_{-n}^k {\bf P}^{N-1}$ models are described by
a gauged linear sigma model with $N$ chiral superfields $x_i$ and one
chiral superfield $z$, charged with respect to a pair of $U(1)$ gauge
symmetries as follows:
\begin{center}
\begin{tabular}{c|cc}
 & $x_i$ & $z$ \\ \hline
$\lambda$ & $1$ & $-n$ \\
$\mu$ & $0$ & $k$ \\
\end{tabular}
\end{center}
Following the prescription of \cite{ronendave,horivafa}, the
`mirror' theory can be obtained from a theory with neutral
chiral superfields $Y_1$, $Y_2$, $\cdots$, $Y_N$, $Y_z$, corresponding
to the $x_i$ and $z$ chiral superfields, and two gauge multiplets
$\Sigma_{\lambda}$, $\Sigma_{\mu}$, with a superpotential\footnote{
We are being sloppy about FI terms and scales.  A meticulous reader will
find it trivially easy to reinsert both in our expressions.}
\begin{displaymath}
\widetilde{W} \: = \: \Sigma_{\lambda} \left( Y_1 \: + \:
\cdots \:  Y_N \: - \: n Y_z \right) \: + \:
\Sigma_{\mu}\left( k Y_z \right) \: + \:
\sum_{i=1}^N \exp(-Y_i) \: + \: \exp(-Y_z)
\end{displaymath}
Integrating out the $Y$'s returns an effective superpotential for
the $\Sigma$'s identical to that calculated from one-loop effects
in \cite{ronendave2}.
Integrating out the $\Sigma$'s gives the constraints\footnote{The two
constraints below would be the D-term equations, if we had included
FI terms above.}
\begin{eqnarray*}
Y_1 \: + \: \cdots \: + \: Y_N \: -\: n Y_z & = & 0 \\
k Y_z & = & 0
\end{eqnarray*}
and an effective superpotential
\begin{equation}  \label{todagnp}
\tilde{W} \: = \: \exp(-Y_1) \: + \: \cdots \: + \:
\exp(-Y_{N-1}) \: + \:
\exp(Y_1) \exp(Y_2) \cdots \exp(Y_{N-1}) \Upsilon^{-n} \: + \:
\Upsilon
\end{equation}
where $\Upsilon = \exp(Y_z)$, and hence is constrained by $\Upsilon^k = 1$,
and where we have eliminated the field $Y_N$ using the other constraint 
equation.  (Note that since the $Y$'s are periodic, the constraint
$k Y_z = 0$ does not eliminate $Y_z$, but rather merely forces it
to define a $k$th root of unity.)

This new theory, with $N-1$ neutral chiral superfields $Y_1$, $\cdots$,
$Y_{N-1}$, together with a ${\bf Z}_k$-valued field $\Upsilon$,
and the effective superpotential above, is our proposed Toda mirror to the 
$G_{-n}^k {\bf P}^{N-1}$ model in the sense that A model correlation functions
of the latter can be calculated as B model correlation functions of the former.
We will show in the next section that correlation functions do indeed match.

In passing, note that for $n=1$ this theory specializes to our
earlier result for the first presentation of the $G_{-1}^k {\bf P}^{N-1}$
model.

\subsubsection{Check of correlation functions of Toda duals}

To check that the Toda theories just described really do correspond
to the gerby ${\bf P}^{N-1}$ models $G_{-n}^k {\bf P}^{N-1}$, we shall
compute the B model correlation functions of the Toda theory,
and check that they match the A model correlation functions of the
gerby ${\bf P}^{N-1}$ models.

In fact, let us back up one step further, and first review how
B model periods are calculated in the Toda theory corresponding
to the ordinary ${\bf C} {\bf P}^{N-1}$ model.

Recall from \cite{vafatoplg} that given a B-twisted Landau-Ginzburg
model with superpotential $W$, if we define $H = \mbox{det}\left(
\partial_i \partial_j W \right)$, then the tree-level correlation
functions can be calculated in the form
\begin{displaymath}
< F_1 \cdots F_n > \: = \: \sum_{dW=0} \frac{F_1 \cdots F_n }{H}
\end{displaymath}
where the $F_i$ are polynomials in the chiral superfields,
{\it i.e.} observables of the B-twisted Landau-Ginzburg model.

The Toda theory corresponding to the ordinary
${\bf P}^{N-1}$ model is a B-twisted Landau-Ginzburg model 
with superpotential
\begin{displaymath}
W \: = \: \exp(Y_1) \: + \: \cdots \: + \:
\exp(Y_{N-1}) \: + \: \exp( -Y_1 - Y_2 - \cdots - Y_{N-1})
\end{displaymath}
Define $\Pi = \exp(-Y_1 -Y_2 - \cdots - Y_{N-1})$;
then the classical vacua are defined by
\begin{displaymath}
\exp(Y_1) \: = \: \cdots \: = \: \exp(Y_{N-1}) \: = \: \Pi
\end{displaymath}
which implies that 
\begin{displaymath}
\left( \exp(Y_i) \right)^N \: = \: 1
\end{displaymath}
for all $i$.
When restricted to classical vacua, it is straightforward to
compute that the determinant $H = N \Pi^{N-1}$.
Thus, if we define $X = \exp(Y_1)$, then for correlation functions we 
compute that
\begin{displaymath}
< X^m > \: = \: \sum \frac{ X^m }{ N X^{N-1} }
\end{displaymath}
using the fact that at classical vacua, $X = \Pi$,
and where the sum runs over $N$th roots of unity.
This expression can only be nonvanishing when $m+1-N$ is
divisible by $N$, thus the only nonvanishing correlation functions are
\begin{displaymath}
< X^{N-1} >, \: < X^{2N-1} >, \:  < X^{3N-1} >, \: \cdots
\end{displaymath}
By comparison, recall that the ordinary ${\bf P}^{N-1}$ model
has quantum cohomology relation $X^N=q$ and classical
correlation function $< X^{N-1} > = 1$,
from which one computes that the nonvanishing correlation functions,
beyond the classical correlation function, are
\begin{eqnarray*}
< X^{2N-1} > & = & q \\
< X^{3N-1} > & = & q^2 
\end{eqnarray*}
and so forth, in agreement with the Toda calculation above.

Now let us turn to the proposed Toda dual to the $G_{-n}^k {\bf P}^{N-1}$
model.  Recall that this Toda theory has $N-1$ neutral chiral superfields $Y_i$,
together with a ${\bf Z}_k$-valued field $\Upsilon$, and the effective
superpotential~(\ref{todagnp}).
It is very straightforward to generalize the methods of \cite{vafatoplg}
to this situation.  Since the ${\bf Z}_k$-valued fields are annihilated by
supersymmetry transformations, and so do not have superpartners,
we can immediately deduce:
\begin{itemize}
\item Each ${\bf Z}_k$-valued field is itself a BRST-invariant observable.
\item As the Hessian factor is derived from F-terms and Yukawa couplings, 
in a theory with
${\bf Z}_k$-valued fields it is calculated only for complex-valued chiral
superfields, omitting the ${\bf Z}_k$-valued fields.
\end{itemize}
Otherwise the form of the calculations in \cite{vafatoplg} are unchanged.
Thus, following \cite{vafatoplg}, tree-level correlators have the form
\begin{displaymath}
< F_1 \cdots F_p > \: = \: \sum_{dW=0} \frac{ F_1 \cdots F_p }{H}
\end{displaymath}
where $H$ is the determinant of the matrix of second derivatives
of the superpotential, taking derivatives corresponding to
complex-valued chiral superfields, and the $F$'s are combinations of
$\exp(Y_i)$ and $\Upsilon$.
The solutions of $dW=0$ are given by
\begin{displaymath}
\exp(-Y_i) \: = \: \exp(Y_1) \exp(Y_2) \cdots \exp(Y_{N-1}) 
\Upsilon^{-n}
\end{displaymath}
for all $i$, and as $\Upsilon$ is ${\bf Z}_k$-valued, all its values
trivially satisfy $dW=0$.
Let us define
\begin{displaymath}
\Pi \: = \: \exp(Y_1) \exp(Y_2) \cdots \exp(Y_{N-1}) \Upsilon^{-n}
\end{displaymath}
so that the classical vacua can be described as
\begin{displaymath}
\exp(-Y_1) \: = \: \exp(-Y_2) \: = \: \cdots \: = \: \exp(-Y_{N-1})
\: = \: \Pi
\end{displaymath}
Then, on the locus of classical vacua, it can be shown that
\begin{displaymath}
H \: = \: N \Pi^{N-1}
\end{displaymath}
Finally, let us define $X = \exp(-Y_1)$, which from the work above
we see satisfies $X^N = \Upsilon^{-n}$ on the locus of classical
vacua.

For reasons already outlined,
correlation functions in this theory can then be described as
\begin{eqnarray*}
< X^m \Upsilon^p > & = & \sum_{dW=0} \frac{X^m \Upsilon^p}{H} \\
& = & \sum_{\Upsilon \in {\bf Z}_k} \sum_{X^N = \Upsilon^{-n}}
\frac{ X^m \Upsilon^p }{N X^{N-1}}
\end{eqnarray*}
This sum will only be nonvanishing when
\begin{displaymath}
m \: - \: N \: + \: 1 \: = \: rN
\end{displaymath}
for some integer $r$ (taking advantage
of $X^N = \Upsilon^{-n}$) such that $p - nr = sk$ for some integer $s$
(taking advantage of $\Upsilon^k = 1$).
For example, if 
\begin{displaymath}
m \: - \: N \: + \: 1 \: = \: rN \: + \: \epsilon
\end{displaymath}
for some integer $\epsilon$ between $0$ and $N$, then the correlation
function has an internal sum of the form $\sum_X X^\epsilon$
over roots $X$ on the classical locus, which vanishes.

In any event, we see that the nonvanishing correlation functions
in this theory are of the form
\begin{displaymath}
< X^{N(r+1)-1} \Upsilon^{nr + sk}>
\end{displaymath}
for integers $r$, $s$, which exactly matches our earlier 
result, equation~(\ref{gnpcorr}) in section~\ref{moregerbes}, 
for A model correlation functions
in the $G_{-n}^k {\bf P}^{N-1}$ model, if we identify $Y \sim \Upsilon^{-1}$.

Let us take a moment to discuss some alternative potential Toda theories
corresponding to the $G_{-1}^k {\bf P}^{N-1}$ model,
and reasons why they are not acceptable.
For example, for the $G_{-1}^k {\bf P}^1$ model, one natural guess
for a corresponding Toda theory would have superpotential
\begin{displaymath}
W \: = \: \exp(k \Theta) \: + \: \exp(-k\Theta)
\end{displaymath}
just as the ordinary ${\bf P}^1$ model has corresponding Toda theory
with superpotential
\begin{displaymath}
W \: = \: \exp(\Theta) \: + \: \exp(-\Theta)
\end{displaymath}
However, the periods of this Toda theory do not reproduce
the A model quantum cohomology calculations discussed earlier.
Define $X = \exp(\Theta)$; then following \cite{vafatoplg}
\begin{displaymath}
<X^m> \: = \: \sum_{dW=0} \frac{ X^m }{ H}
\end{displaymath}
In the present case,
\begin{displaymath}
H \: = \: k^2 \left( \exp(k \Theta) \: + \: \exp(-k\Theta) \right)
\: = \: k^2\left( X^k \: + \: X^{-k} \right)
\end{displaymath}
and the classical vacua are defined by $X^k = X^{-k}$ or
$X^{2k}=1$.  Thus, the classical vacua are the $2k$th roots
of unity, so we do at least have as many classical vacua as expected.
The B model correlation functions, however, are more problematic:
\begin{displaymath}
<X^m> \: \propto \: \sum X^{m-k}
\end{displaymath}
where the sum is over $2k$th roots of unity,
and hence is only nonvanishing when $m-k$ is a multiple of $2k$,
which implies that the nonzero correlation functions are
\begin{displaymath}
< X^{k} >, \: < X^{3k} > , \: < X^{5k} >, \cdots
\end{displaymath}
However, previously we have calculated for this model that the
nonvanishing correlation functions are given by
\begin{displaymath}
<X> = 1, \: <X^{2k+1}> = q, \: < X^{4k+1}> = q^2, \: \cdots
\end{displaymath}
which does not match the result above,
except in the trivial case $k=1$.

Furthermore, the obvious extension of the alternate proposal above to
$G_{-1}^k {\bf P}^{N-1}$, namely the Toda theory with superpotential
\begin{displaymath}
W \: = \: \exp(kY_1) \: + \: \cdots \: + \: \exp(kY_{N-1}) \: + \:
\exp(-kY_1 -kY_2 - \: \cdots \: - kY_{N-1})
\end{displaymath}
not only does not have matching correlation functions, but does not
even have the expected number of classical vacua (it has
$kN(N-1)$ instead of $kN$).

\subsection{Mirrors to gerbes on the quintic}

Let us begin our discussion of mirrors to gerbes over the quintic
by appealing to a simple minimal-model argument at the Fermat 
Landau-Ginzburg point
in the moduli space.  As noticed previously in section~\ref{lgorb},
the Landau-Ginzburg orbifold corresponding to the quintic hypersurface in
$G_{-1}^k {\bf P}^{N-1}$ is a noneffectively-acting ${\bf Z}_{5k}$ orbifold.
Thus, at the Fermat point, the theory can be constructed
as a ${\bf Z}_{5k}$ orbifold of a tensor product of five minimal models,
call them each $M$:
\begin{displaymath}
\left[ \frac{ M \otimes M \otimes \cdots \otimes M }{ {\bf Z}_{5k} } \right]
\end{displaymath}
Now, by ordinary mirror symmetry for the minimal models,
we can replace each $M$ by the ${\bf Z}_5$ orbifold
$[M/{\bf Z}_5]$, leaving us with a ${\bf Z}_{5k}$ orbifold
of a tensor product of five copies of $[M / {\bf Z}_5]$.
As shown in \cite{nr}, a noneffective ${\bf Z}_{kn}$ orbifold
of a ${\bf Z}_n$ orbifold is a trivially-acting ${\bf Z}_k$ orbifold.
Thus, in the present case we are left with a trivially-acting
${\bf Z}_k$ orbifold of a ${\bf Z}_5^4$ orbifold of a tensor product
of five minimal models, {\it i.e.} a trivially-acting ${\bf Z}_k$ orbifold
of the mirror to the standard Landau-Ginzburg Fermat model.

In light of the result \cite{nr} that the mirror to a trivial
gerbe is a trivial gerbe over the mirror, the calculation above implies
that, at the Fermat Landau-Ginzburg point, the conformal field theory
of the ${\bf Z}_{5k}$ orbifold is the same as a 
${\bf Z}_5 \times {\bf Z}_k$ orbifold.  Note that this statement is consistent
with partition functions:  since ${\bf Z}_{5k}$ is abelian,
the $g$-loop partition function of the ${\bf Z}_{5k}$ orbifold is the
same as that of the ${\bf Z}_5 \times {\bf Z}_k$ orbifold, for all $g$.

Let us now formally compute the mirror of the gerby quintic using the
methods of \cite{ronendave,horivafa,agvafa}.  Since this is a hypersurface
in a toric variety, and not a toric variety, we shall use the trick of
\cite{agvafa} of replacing the A model on such a hypersurface with the
A model on a toric supervariety, which can be dualized using the methods
of \cite{ronendave,horivafa} to an effective Landau-Ginzburg theory whose
B model correlation functions should duplicate the original A model
correlation functions.  (Unfortunately, this is a many-to-one map,
as complex structure data is lost on the A side, corresponding to the
fact that the dual is fixed at one point in K\"ahler moduli space.
Nevertheless, this will give good insight into the structure of the 
mirrors.) 

Following \cite{horivafa}, let us first compute the mirror of the
ambient gauged linear sigma model with no superpotential.
The mirror can be described starting from a theory with neutral
chiral superfields $Y_1$, $\cdots$, $Y_5$, $Y_p$, and $\Sigma$,
and effective
superpotential
\begin{displaymath}
\widetilde{W} \: = \: \Sigma \left( -5k Y_p \: + \: k Y_1 \: + \:
\cdots \: + \: k Y_5  -  k r \right) \: + \: \sum_i \exp(-Y_i) \: + \:
\exp(-Y_p)
\end{displaymath}
where $r$ is proportional to the Fayet-Iliopoulos term.
(We have included it here, with a suitable normalization,
to make the result cleaner.)
For simplicity, we will assume that $k$ is not divisible by $5$.
Integrating out $\Sigma$ gives us the constraint
\begin{displaymath}
k Y_p \: = \: - \frac{1}{5} k \left( Y_1 \: + \: \cdots \: + \: Y_5 \: - \: r
\right)
\end{displaymath}
Since the fields $Y$ are periodic, if we define $X_i = \exp(-Y_i/5)$,
then we have that
\begin{displaymath}
\exp(-Y_p) \: = \: \Upsilon \exp(-r/5) X_1 X_2 X_3 X_4 X_5
\end{displaymath}
where $\Upsilon$ is a $k$th root of unity, whose possible values must
be summed over in the path integral measure.
Furthermore, because $X_i = \exp(-Y_i/5)$, we have identifications
\begin{displaymath}
X_i \: \sim \: \omega_i X_i
\end{displaymath}
where $\omega_i$ is a fifth root of unity.
Using the assumption that $5$ does not divide $k$,
the relation between $Y_p$ and the $Y_i$ tells us that
$\prod_i \omega_i = 1$, so in fact we must orbifold by ${\bf Z}_5^4$
rather than ${\bf Z}_5^5$.
Thus, we have an effective Landau-Ginzburg model of the mirror defined
by the effective superpotential
\begin{displaymath}
\widetilde{W} \: = \: X_1^5 \: + \: \cdots \:  + \: X_5^5 \: + \:
\Upsilon \exp(-r/5) X_1 X_2 X_3 X_4 X_5
\end{displaymath}
and the ${\bf Z}_5^4$ orbifold action described previously,
where $\Upsilon$ is a $k$th root of unity, and the path integral 
measure must sum over values of $\Upsilon$.
So far we have only computed the mirror of the gauged linear sigma
model with vanishing superpotential.  In this theory, the fundamental
fields\footnote{As a result, for example, the untwisted part
of the chiral ring in the dual to the
theory with no superpotential is given by the invariant part of
${\bf C}[Y_1, Y_2, \cdots, Y_5]/(\partial W)$ with $W$ of the form above,
an infinite-dimensional vector space, reflecting the noncompactness
of the original theory.  Turning on the superpotential in the original theory
yields a dual with chiral ring untwisted sector given by the invariant part of
${\bf C}[X_1, X_2, \cdots, X_5]/(\partial W)$,
a finite-dimensional vector space.  We would like to thank A.~Adams for
discussions of these matters.} 
over which the path integral measure sums are the $Y$'s, not the $X$'s.
It is easy to check that
turning on a superpotential, and proceeding as in \cite{agvafa},
has the ultimate effect only of changing the fundamental fields from the $Y$'s
to the $X$'s.  Otherwise, just as for the quintic in ordinary ${\bf P}^4$,
the form of the effective Landau-Ginzburg theory is the same between the
two cases.
In particular, the effective Landau-Ginzburg theory dual to the
A model on the gerby quintic is defined by the superpotential above,
with the field $\Upsilon$ taking values in roots of unity.

At the Fermat point, where the $\prod_i X_i$ term decouples,
the theory reduces to a ${\bf Z}_5^4$ orbifold of
a product of five minimal models, tensored with the theory of
the ${\bf Z}_k$-valued field $\Upsilon$.
However, as discussed in \cite{nr}, tensoring with the trivial
theory of a ${\bf Z}_k$-valued field is equivalent to orbifolding by
a trivially-acting ${\bf Z}_k$, so we recover the same result
at the Fermat point that we derived earlier using nothing more than
standard facts about minimal models.

Away from the Fermat point, we see something interesting.
The ${\bf Z}_k$-valued field $\Upsilon$, which at Fermat merely
realizes a trivially-acting ${\bf Z}_k$ orbifold, now mixes
nontrivially with superpotential terms.
Recall that the same sort of structure appeared from
an independent line of thought in \cite{nr} when discussing
the interpretation of moduli in twisted sectors associated to 
trivially-acting group elements.
In hindsight, this should not be surprising -- both here and
in \cite{nr}, we are ultimately studying the same sorts of problems, namely
deformations of stringy moduli appearing in gerbes.
Thus, we should not be surprised to see the same structures emerging
in both cases.

\subsection{Gerby minimal models}

Let us take just a moment to outline some easy results concerning
mirror symmetry and noneffective orbifolds of minimal models.

Consider the $A_{d-1}$ minimal model, which arises as the IR fixed point
of the LG model of a single chiral superfield $\Phi$ with superpotential
$\Phi^d$.  It is well known that
that model is mirror to its orbifold by ${\bf Z}_d$,
acting as $\Phi \mapsto \exp(2 \pi i / d) \Phi$.
We can generate an easy example of gerby mirrors in minimal models
using the result in \cite{nr} that for any CFT ${\cal C}$,
if we define ${\cal C}'$ to be the orbifold $[ {\bf C}/{\bf Z}_d]$,
then the orbifold $[ {\bf C}' / {\bf Z}_{dk} ]$ (where ${\bf Z}_d$
is the quantum symmetry and ${\bf Z}_k$ acts trivially)
is isomorphic to the orbifold $[ {\cal C} / {\bf Z}_k ]$.
Applied to the minimal model above,
this means that the ${\bf Z}_{dk}$ orbifold of the $A_{d-1}$ minimal
model (where the ${\bf Z}_{dk}$ acts by projecting to ${\bf Z}_d$,
acting in the usual way, with trivial ${\bf Z}_k$ kernel)
is isomorphic to the trivial ${\bf Z}_k$ orbifold of the same
minimal model.

Furthermore, from the argument in \cite{nr} that trivial gerbes are mirror,
one expects that the trivial $[A_{d-1}/{\bf Z}_k]$ orbifold should be 
mirror to $[ [ A_{d-1}/{\bf Z}_d ] / {\bf Z}_k ] \cong
[ A_{d-1}/ \left( {\bf Z}_d \oplus {\bf Z}_k \right) ]$,
hence the ${\bf Z}_{kd}$ and ${\bf Z}_d \oplus {\bf Z}_k$ orbifolds
of the $A_{d-1}$ minimal model should be isomorphic.
It is trivial to check that spectra and $g$-loop partition functions match
between two such orbifolds, hence the result seems very plausible.

\subsection{Analogue of Batyrev's mirror conjecture for stacks}

In order to find an analogue of Batyrev's mirror conjecture 
\cite{batyrev} one faces the following basic puzzle:  since a toric
stack is described in terms of a fan plus finite group data associated
to each edge (see the appendix for more information), 
in order for reflexive polyhedra to encode mirrors,
Newton polygons must somehow be decorated with the same sort of finite
group data that appears in toric stacks. 

This puzzle is related to the puzzle posed by deformation theory of stacks:
the only mathematical deformations of a stack are those in the untwisted
sector.  Twist field moduli do not have a purely mathematical understanding.

We have seen how explicit mirror constructions solve the deformation
theory puzzle, by forcing us to introduce a new class of abstract CFT's,
based on Landau-Ginzburg orbifolds, in which the superpotential is deformed
by finite-group-valued fields, corresponding to giving a vev to a 
twist field in a noneffective orbifold.

The solution to the puzzle in generalizing Batyrev's conjecture to toric
stacks is the same:  we need to consider the same general class of
abstract CFT's, in which superpotentials naturally include monomial
terms with finite group factors.

It is now straightforward to generalize Batyrev's conjecture.
Recall that in its ordinary form, given a Calabi-Yau hypersurface in a toric
variety, one first constructs the Newton polygon of 
possible monomial contributions
to the hypersurface, and another polygon with faces over the cone of the
fan describing the ambient toric variety.  (One typically scales so that
each polyhedron, on an integral lattice, has exactly one interior point.)
So long as these polyhedra are
reflexive, the mirror is obtained by exchanging these polyhedra:
the mirror ambient toric variety has fan defined by the Newton polyhedron
of the original, and the monomials in a Calabi-Yau hypersurface are described
by the polyhedron over the fan of the original ambient toric variety.
(See \cite{candmirrev} for a very readable discussion.)

To generalize this procedure to stacks, we proceed as follows.
The stacky fan of the mirror ambient toric variety is obtained by
using the above construction to get the basic fan, with finite-group data
obtained by taking the difference of finite-group factors between
the monomial corresponding to that corner of the polyhedron and the monomial
corresponding to the interior point.  The Newton polyhedron of the mirror
theory is obtained from the same procedure as for the standard construction,
with finite-group-valued-field-factors obtained from the generators decorating
each edge of the stacky fan.

Let us work through a simple example, to make this process clearer.
Let us first review Batyrev's mirror construction for elliptic curves,
built as degree three hypersurfaces in ${\bf P}^2$,
and then we shall describe how to generalize this process to describe
the mirror of the gerbe over an elliptic curve obtained by restricting
the $G_{-1}^k {\bf P}^2$ model to the hypersurface.

\begin{figure}
\centerline{\psfig{file=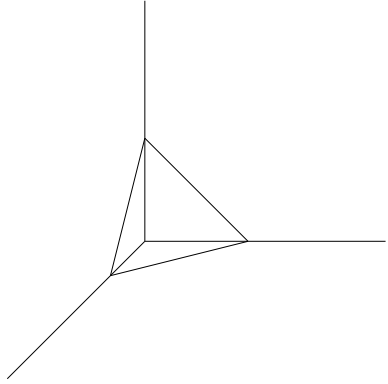,width=2.0in}}
\caption{\label{p2m1} The fan for ${\bf P}^2$, with the mirror Newton
polyhedron shown.}
\end{figure}

\begin{figure}
\centerline{\psfig{file=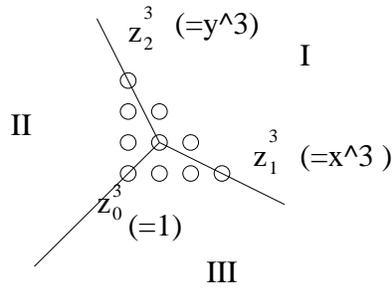,width=2.0in}}
\caption{\label{p2m2} The Newton polyhedron for degree three hypersurfaces
in ${\bf P}^2$.  The three corner points, corresponding to
cubes of the three homogeneous coordinates, are labelled.
The fan of the mirror ambient toric variety, the fan over the 
faces of the Newton polyhedron, is shown.}
\end{figure}

In figure~\ref{p2m1} we have shown the fan for the toric variety
${\bf P}^2$.  In figure~\ref{p2m2} we have shown the Newton polyhedron
for degree three hypersurfaces in ${\bf P}^2$.

The fan of the mirror ambient toric variety is obtained as the
fan through the faces of the Newton polyhedron, and is shown in
figure~\ref{p2m2}.  It is straightforward to compute that this
fan describes the toric variety ${\bf P}^2/{\bf Z}_3$, where if
we denote the homogeneous coordinates on ${\bf P}^2$ by
$[z_0, z_1, z_2]$, and let $\xi$ generate the third roots of unity,
then the generator of the ${\bf Z}_3$ acts as
\begin{displaymath}
\left[ z_0, z_1, z_2 \right] \: \mapsto \:
\left[ z_0, \xi z_1, \xi^2 z_2 \right]
\end{displaymath}
In a little more detail, if we let the three cones in the fan
be denoted $I$, $II$, $III$, where $I$ is the cone covering the
first quadrant and the others proceed counterclockwise,
then the coordinate patch corresponding to cone $I$ is given by
\begin{displaymath}
\mbox{Spec } {\bf C} \left[ x^2 y, xy^2, xy \right]
\end{displaymath}
the coordinate patch corresponding to cone $II$ is given by
\begin{displaymath}
\mbox{Spec } {\bf C} \left[ x^{-2} y^{-1}, x^{-1} y, x^{-1} \right]
\end{displaymath}
and the coordinate patch corresponding to cone $III$ is given by
\begin{displaymath}
\mbox{Spec } {\bf C}\left[ x y^{-1}, x^{-1} y^{-2}, y^{-1} \right]
\end{displaymath}
We can relate this description to the homogeneous coordinates on ${\bf P}^2$
by identifying
\begin{displaymath}
x \: \sim \: \frac{z_1^2}{z_0 z_2}, \: \: \:
y \: \sim \: \frac{z_2^2}{z_0 z_1}
\end{displaymath}

\begin{figure}
\centerline{\psfig{file=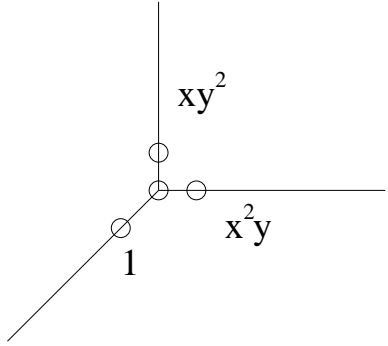,width=2.0in}}
\caption{\label{p2m3} Newton polyhedron for degree three hypersurfaces
in ${\bf P}^2/{\bf Z}_3$, with fan over the faces shown.}
\end{figure}

The Newton polyhedron for degree three hypersurfaces in ${\bf P}^2/{\bf Z}_3$
has monomials
\begin{displaymath}
\{ z_0^3, z_1^3, z_2^3, z_0 z_1 z_2 \} \: = \:
\{ 1, x^2 y, xy^2, xy \}
\end{displaymath}
and is shown in figure~\ref{p2m3}.

Note that the fan shown in figure~\ref{p2m3}, constructed from the 
mirror Newton polyhedron, is the same as the fan of the original
${\bf P}^2$, so that applying the mirror construction twice returns to
the original toric variety and hypersurface.

Now that we have seen how Batyrev's construction works for the case
of elliptic curves in ${\bf P}^2$, let us turn to a simple example
of a toric stack.  Consider the ${\bf Z}_k$ gerbe over ${\bf P}^2$
with characteristic class $-1 \mbox{ mod } k$.  As described in more
detail in the appendix, the stacky fan has underlying fan given by
the same fan as ${\bf P}^2$, and to the edges we associate generators
of ${\bf Z}^2 \oplus {\bf Z}_k$, given by
\begin{equation}   \label{p2gerbe}
\begin{array}{c}
(1,0,0) \\
(0,1,0) \\
(-1,-1,1)
\end{array}
\end{equation}
If we consider gerbes over ordinary elliptic curves, so that the
possible monomials do not have any finite-group factors,
then the mirror ambient toric stack has fan given by that for
${\bf P}^2/{\bf Z}_3$, and to the edges we associate generators
of ${\bf Z}^2 \oplus {\bf Z}_k$ given by
\begin{equation}
\begin{array}{c}
(-1,2,0) \\
(2,-1,0) \\
(-1,-1,0)
\end{array}
\end{equation}
Since there are no finite-group factors in the monomials, the generators
of ${\bf Z}^2 \oplus {\bf Z}_k$ do not extend into ${\bf Z}_k$.
The Newton polygon of the mirror is constructed from the polygon
over the fan of the ambient toric stack.  
The possible monomials are weighted by finite-group data, in a fashion
determined by the generators~(\ref{p2gerbe}) of the original toric stack.
Specifically, we have monomial terms
\begin{displaymath}
\{ \Upsilon z_0^3, z_1^3, z_2^3, z_0 z_1 z_2 \}
\end{displaymath}
in the superpotential of the mirror Landau-Ginzburg theory, 
where $\Upsilon$ is a ${\bf Z}_k$-valued
field, which appears in the $z_0^3$ monomial because the generator
associated to the corresponding edge of the toric fan extends into
${\bf Z}_k$.
If the generators associated to the edges had been
\begin{displaymath}
\begin{array}{c}
(1,0,a) \\
(0,1,b) \\
(-1,-1,c)
\end{array}
\end{displaymath}
then the monomial terms in the superpotential of the mirror Landau-Ginzburg 
model would have been
\begin{displaymath}
\{ \Upsilon^c z_0^3, \Upsilon^a z_1^3, \Upsilon^b z_2^3, z_0 z_1 z_2 \}
\end{displaymath}
Similarly, if there were additional finite-group factors in the defining
data of the stacky fan, there would be additional finite-group-valued
fields appearing in the mirror Landau-Ginzburg superpotentials.

Also note that if we work with a stacky fan with ${\bf Z}_k$ factors
but no generator extends into those factors, and the superpotential
has only ordinary monomials -- no finite-group-valued field factors -- 
then the mirror of the stack is the mirror Calabi-Yau, in the ordinary
sense, in a stacky fan with ${\bf Z}_k$ factors.  This can be interpreted
as the statement that the mirror of the trivial ${\bf Z}_k$-gerbe
over a Calabi-Yau is the trivial ${\bf Z}_k$-gerbe over the mirror
Calabi-Yau, agreeing with easy general statements on mirrors of
trivial gerbes described in \cite{tonyme,nr}.

In passing, note that the monomial-divisor mirror map now immediately
generalizes to toric stacks, as we have implicitly constructed a map
between generators of the K\"ahler-like deformations of the
theories with finite-group-valued fields, and the complex-structure-like
deformations, in the same form as the original monomial-divisor mirror map.

In order to properly understand whether this proposed generalization
of Batyrev's mirror conjecture is physically correct, we would need
to further pursue properties of these new abstract CFT's constructed
from Landau-Ginzburg orbifolds with finite-group-valued fields.
We will leave such a study for future work, and for the purposes of this
paper content ourselves with having merely elucidated the conjecture.

\section{Conclusions}

In this paper we have discussed gauged linear sigma models with nonminimal
$U(1)$ charges.  Such theories, which describe quotients by 
noneffectively-acting ${\bf C}^{\times}$'s, are physically distinct from
their counterparts with minimal charges (as we saw in the analogous case
of finite noneffectively-acting groups in \cite{nr}),
and can be understood mathematically as describing toric stacks.

Every stack has a presentation of the form $[X/G]$, to which one associates
a $G$-gauged sigma model on $X$.  However, such presentations are not unique.
We have argued in \cite{tonyme} and seen further examples here showing
that stacks classify universality classes of worldsheet RG flow in
gauged sigma models.  Curiously, this seems to be true not only for 
Calabi-Yau stacks, but also for the A model on more general stacks,
as we have seen in examples here.

We have discussed quantum cohomology of gauged linear sigma models
corresponding to toric stacks, as well as massless spectra of
Landau-Ginzburg models associated to gerby hypersurfaces,
and seen how the results in all such cases are consistent with
the general conjecture for massless spectra in gauged sigma models
presented in \cite{tonyme}, namely that the massless spectrum is the
de Rham cohomology of the associated inertia stack.

Finally, and perhaps most importantly, we have discussed mirror symmetry
for stacks, and seen how fields valued in roots of unity play an
important role, in Toda duals to toric stacks, in the mirror to the
gerby quintic, and more generally in formulating Batyrev's mirror conjecture
for hypersurfaces in toric stacks.  
Previously in \cite{nr,tonyme} we have seen such
fields appearing when understanding deformations along marginal operators
in certain twisted sectors of noneffective orbifolds, and used them to
give a very explicit understanding of the CFT's one obtains by such
deformations.  Here, although fields valued in roots of unity are derived
from completely independent lines of reasoning, the root cause of their
appearance is the same:  their appearance is again needed to understand
moduli spaces of theories containing noneffectively-gauged sigma models
at certain points.

\section{Acknowledgements}

We would like to thank A.~Adams, L.~Chen, J.~Distler, K.~Hori, S.~Katz,
J.~McGreevy, and R.~Plesser
for useful conversations, A.~Greenspoon for proofreading,
the Aspen Center for Physics
for hospitality while this work was under development,
and the UPenn Math-Physics group for the excellent conditions for collaboration
it provided during several stages of this work.
T.P. was partially supported by NSF grants DMS 0403884 and
FRG 0139799.

\appendix

\section{Toric stacks}

The paper \cite{bcs} describes a construction of ``toric stacks,''
which are Deligne-Mumford stacks over toric varieties.

Toric stacks are described as follows.
Begin with a fan describing a toric variety.
(The toric stack will naturally live over the toric variety;
technically, the toric variety is known as the ``moduli space'' of the stack,
though in this context the term is a holdover from the historical development
of stacks, and does not refer
to any deformation theory.)
Enlarge the lattice $N$ to include some finite group factors,
so that if for the original toric variety, $N = {\bf Z}^d$ for some $d$,
then $N$ is now of the form
\begin{displaymath}
{\bf Z}^d \oplus {\bf Z}_{q_1} \oplus \cdots \oplus {\bf Z}_{q_r}
\end{displaymath}
For ordinary toric varieties, each edge in the fan describes a map
${\bf Z} \rightarrow N$, defined by a generator of that edge.
We do the same here -- to each edge we associate an element of
the abelian group $N$ above.  The freely-generated part of the image
corresponds to the coordinates of a generator of the edge,
and the finite-group factors will determine the stack structure.
The combination of an ordinary fan, $N$, and the maps from edges to
$N$ forms what the authors of \cite{bcs} call a {\it stacky fan}.

We can determine ${\bf C}^{\times}$ charges in a 
gauged-linear-sigma-model-style presentation
as follows.
Let $n$ be the number of edges in the fan, and $d$ the dimension
of the freely-generated part of $N$.  Define a $(d+r) \times r$ matrix
$Q$ to have $0$'s in its first $d$ rows, then $q_i$'s on the diagonal
in the last $r$ rows, giving a matrix of the form
\begin{displaymath}
Q \: = \:
\left[ \begin{array}{cccc} 
       0 & 0 & \cdots & 0 \\
         &   & \cdots &   \\
       0 & 0 & \cdots & 0 \\
       q_1 & \cdots & 0 & 0 \\
       0 & q_2 & \cdots & 0 \\
       0 & \cdots & q_{r-1} & 0 \\
       0 & 0 & \cdots & q_{r}
       \end{array}  
\right].
\end{displaymath}
Define a $(d+r) \times n$ matrix $P$ to have, in each column,
the image in $N$ of the corresponding edge.
Then, the gauged-linear-sigma-model charges are defined by the kernel of the
$(d+r)\times(n+r)$ matrix $[P Q]$, a result in close analogy with the way
gauged-linear-sigma-model charges are computed from ordinary fans for
ordinary toric varieties.
Technically there can sometimes be additional finite group actions
beyond the ${\bf C}^{\times}$'s; we shall return to this matter after
describing a few basic examples.

Our first example is also example~2.1 in \cite{bcs}.
The underlying toric variety is ${\bf P}^1$, with its
trivial fan, and take $N = {\bf Z} \oplus {\bf Z}_2$.
Since the toric variety is one-dimensional, $d=1$,
and since there is only one finite group factor $r=1$,
and in fact the matrix $Q$ is given by
\begin{displaymath}
\left[ \begin{array}{c}
       0 \\ 2 \end{array} \right]
\end{displaymath}
To the first edge we associate $(2,1) \in N$,
and to the second edge we associate $(-3,0) \in N$.
Thus, the matrix $P$ is given by
\begin{displaymath}
\left[ \begin{array}{cc}
       2 & -3 \\
       1 & 0 
       \end{array} \right]
\end{displaymath}
The gauged-linear-sigma-model charges are computed as the solutions
$a$, $b$, $c$ of the equation
\begin{displaymath}
\left[ \begin{array}{ccc}
       2 & -3 & 0 \\
       1 & 0 & 2 
       \end{array} \right]
\left[ \begin{array}{c}
       a \\ b \\ c \end{array} \right]
\: = \: 0
\end{displaymath}
The possible solutions are generated by
\begin{displaymath}
\left[ \begin{array}{ccc}
       a & b & c
      \end{array} \right] 
\: = \:
\left[ \begin{array}{ccc}
       6 & 4 & -3
      \end{array} \right] 
\end{displaymath}
We discard the value of $c$, and then the resulting stack can 
be realized by a gauged linear sigma model acting on two chiral superfields,
with a single $U(1)$, under which the chiral superfields have charges
$6$, $4$.  Truncating $c$ may seem somewhat artificial, but has a natural
mathematical interpretation, which we shall return to later.

Suppose we modify the example above slightly,
using the same underlying toric fan and the same $N$,
but we modify the maps from the edges into $N$ so that they produce
no torsion, {\it i.e.}, to the edges we associate $(2,0)$ and $(-3,0)$.
Now when we compute the gauged linear sigma model charges as the kernel
of 
\begin{displaymath}
\left[ \begin{array}{ccc}
       2 & -3 & 0 \\
       0 & 0 & 2 
       \end{array} \right]
\end{displaymath}
we find that the charges are $3$, $2$.
Adding the torsion forces us to use non-minimal charges.

Technically, the example above describes more than just
the weighted projective space $W{\bf P}_{3,2}$;
it is actually the stack $W {\bf P}_{3,2} \times B {\bf Z}_2$.
More generally, putting all $0$'s in a row of the $P$ matrix will result
in a $B {\bf Z}_r$ factor.

Let us take a moment to understand the details of the construction in 
\cite{bcs} 
a little better, which will allow us to see the origin of these finite
group factors.
The paper \cite{bcs} defines a group $DG(\beta)$ to be 
$\left( {\bf Z}^{n+r} \right)^*$ modulo the image of the matrix
$[PQ]^T$.
The same reference also defines a map $\beta^{\vee}$, which turns out
to be given by the composition of the inclusion 
\begin{displaymath}
\left( {\bf Z}^n \right)^*
\: \hookrightarrow \: \left( {\bf Z}^{n+r} \right)^*
\end{displaymath}
and the
projection 
\begin{displaymath}
\left( {\bf Z}^{n+r} \right)^* \: \longrightarrow \: DG(\beta).
\end{displaymath}
This projection can be accomplished by contracting the elements of
$\left( {\bf Z}^{n+r} \right)^{\vee}$ with elements of the kernel
of $[P Q]$, the same kernel that defines the gauged linear sigma model
charges.
(The map $\beta$ is the assignment of an element of $N$ to each
edge of the fan, thus, a map ${\bf Z}^n \rightarrow N$.)
To compute finite group factors and their actions, we need to compute
$DG(\beta)$ and $\beta^{\vee}$ in each example.
The group one quotients by, is $G = \mbox{Hom}(DG(\beta), {\bf C}^{\times})$.
If $DG(\beta) = {\bf Z}^k$ for some $k$, then $G = \left( {\bf C}^{\times}
\right)^k$, and we have an ordinary gauged linear sigma model.
If $DG(\beta)$ contains finite group factors, then we must quotient by
more than merely ${\bf C}^{\times}$'s to obtain the toric stack.
Furthermore, the action of $G$ on the homogeneous coordinates
is defined by the map $\mbox{Hom}(\beta^{\vee}, {\bf C}^{\times})$.

In the first example we discussed above,
$DG(\beta) = {\bf Z}$.  The map $\beta^{\vee}$ is the composition of
the inclusion
\begin{displaymath}
\left( {\bf Z}^2 \right)^* \: \longrightarrow \: \left( {\bf Z}^3 \right)^*:
\: \left[ \begin{array}{c} a \\ b \end{array} \right]
\: \mapsto \:
\left[ \begin{array}{c} a \\ b \\ 0 \end{array} \right]
\end{displaymath}
and the projection
\begin{displaymath}
\left( {\bf Z}^3 \right)^* \: \longrightarrow \: {\bf Z}: \:
\left[ \begin{array}{c} a \\ b \\ c \end{array} \right]
\: \mapsto \:
\left[ 6 4 -3 \right] \left[ \begin{array}{c} a \\ b \\ c \end{array}
\right] \: = \: 6a \: + \: 4b \: - \: 3c
\end{displaymath}
More simply, the map $\beta^{\vee}$ is the map
\begin{displaymath}
\left( {\bf Z}^2 \right)^{\vee} \: \longrightarrow \: {\bf Z}: \:
\left[ \begin{array}{c} a \\ b \end{array} \right] \: \mapsto \:
6a \: + \: 4b
\end{displaymath}
Since $DG(\beta) = {\bf Z}$, $G = {\bf C}^{\times}$,
and the action of $G$ on the homogeneous coordinates is given by
$\mbox{Hom}\left( \beta^{\vee}, {\bf C}^{\times} \right)$ or more
simply,
\begin{displaymath}
(x,y) \: \mapsto \: \left( \lambda^6 x, \lambda^4 y \right)
\end{displaymath}
More generally, it is clear from the form of the first inclusion map
that the ${\bf C}^{\times}$ charges of the homogeneous coordinates
will be obtained by truncating the kernel vectors.

In the second example we discussed above,
$DG(\beta) = {\bf Z} \oplus {\bf Z}_2$.
The map $\beta^{\vee}$ is the composition of the inclusion
above, plus the projection
\begin{displaymath}
\left( {\bf Z}^3 \right)^* \: \longrightarrow \: {\bf Z} \oplus {\bf Z}_2: \:
\left[ \begin{array}{c} a \\ b \\ c \end{array} \right]
\: \mapsto \:
\left[ 3 2 0 \right] \left[ \begin{array}{c} a \\ b \\ c \end{array}
\right] \: = \: 3a \: + \: 2b 
\end{displaymath}
In other words, the map $\beta^{\vee}$ is the map
\begin{displaymath}
\left( {\bf Z}^2 \right)^{\vee} \: \longrightarrow \: {\bf Z} \oplus {\bf Z}_2:
\:
\left[ \begin{array}{c} a \\ b \end{array} \right] \: \mapsto \:
\left( 3a \: + \: 2b \right) \oplus 0
\end{displaymath}
Here, $G = {\bf C}^{\times} \oplus {\bf Z}_2$,
and dualizing $\beta^{\vee}$, we find that the action of ${\bf C}^{\times}$
on the homogeneous coordinates is given by
\begin{displaymath}
(x,y) \: \mapsto \: \left( \lambda^3 x, \lambda^2 y \right)
\end{displaymath}
and the ${\bf Z}_2$ acts trivially.
Thus, here the toric stack is $W {\bf P}_{2,3} \times B {\bf Z}_2$.

We can also realize the gerbe of section~\ref{basicpngerbe} in this
language.  Recall that was a ${\bf Z}_k$ gerbe over ${\bf P}^{N-1}$,
which generated all of the ${\bf Z}_k$ gerbes on ${\bf P}^{N-1}$.
Since it is a ${\bf Z}_k$ gerbe, the matrix $Q$ is an $N$-element
column vector of the form
\begin{displaymath}
Q \: = \: \left[ \begin{array}{c}
                 0 \\ 0 \\ \cdots \\ 0 \\ k \end{array} \right]
\end{displaymath}
and if we take the matrix $P$ to be given by
\begin{displaymath}
P \: = \: \left[ \begin{array}{ccccc}
                 1 & 0 & \cdots & 0 & -1 \\
                 0 & 1 & \cdots & 0 & -1 \\
                   &   & \cdots &   &  \\
                 0 & 0 & \cdots & 1 & -1 \\
                 0 & 0 & \cdots & 0 & 1 \end{array} \right]
\end{displaymath}
then column vectors that solve the equation
\begin{displaymath}
[ P Q ] \left[ \begin{array}{c}
               a_1 \\ \cdots \\ a_N \\ b \end{array} \right]
\: = \: 0
\end{displaymath}
are generated by
\begin{displaymath}
\left[ \begin{array}{c}
       a_1 \\ a_2 \\ \cdots \\ a_N \\ b \end{array} \right] \: = \:
\left[ \begin{array}{c}
       k \\ k \\ \cdots \\ k  \\ -1 \end{array} \right]
\end{displaymath}
from which we see the gauged-linear-sigma-model-type description
in terms of $N$ chiral superfields each of charge $k$.
It is straightforward to check that $DG(\beta) = {\bf Z}$,
so there are no extra finite-group factors here.

We do not need to add any finite groups to the $N$ lattice in
order to obtain a stack that is not a space; we can get stacks
from the data defining an ordinary toric variety, if it is singular.
For a smooth toric variety, if we reinterpret the same data
defining the variety as a stack, then the resulting stack will be the
same as the original toric variety.
However, if we take a singular toric variety, such as the
quotient space ${\bf C}^2/{\bf Z}_2$, and interpret the 
fan data as a toric stack, then the resulting stack is
not the space ${\bf C}^2/{\bf Z}_2$, but rather is the stack
$[ {\bf C}^2/{\bf Z}_2]$.
Similarly, weighted projective stacks are described as toric stacks
using the same fan and $N$ lattice as for the corresponding toric
varieties, but merely reinterpreted as stacks.

This is how our conjectured extension of Batyrev's mirror proposal 
to stacks reproduces mirror symmetry for spaces.
Interpreted as stacks, toric varieties such as the
quotient space ${\bf P}^2/{\bf Z}_3$ are replaced by toric stacks
$[ {\bf P}^2/{\bf Z}_3]$, reproducing orbifold mirrors.

\end{document}